\documentclass[11pt,a4paper]{article}
\pdfoutput=1

\usepackage{jheppub}
\usepackage{latexsym}
\usepackage{multirow}
\usepackage{color}
\usepackage[usenames,dvipsnames,table]{xcolor}
\usepackage{graphicx}
\usepackage{epsfig}  
\usepackage{epsf}    
\usepackage{dcolumn}
\usepackage{bm}
\usepackage{dcolumn}
\usepackage{textcomp}
\usepackage{float}
\usepackage{subcaption}
\usepackage{hypcap}
\usepackage[]{hyperref}
\usepackage{makecell}
\usepackage{color}
\usepackage{pifont}
\usepackage{appendix}
\usepackage{amsmath}
\usepackage{multirow,bigdelim}  
\usepackage{lineno}
\usepackage[normalem]{ulem}
\usepackage{diagbox}
\usepackage{adjustbox}

\usepackage{bm}
\usepackage{orcidlink}
\usepackage{wasysym}
\usepackage{mathrsfs}
\usepackage{caption}
\hypersetup{
 unicode=false,          
 pdftoolbar=true,        
 pdfmenubar=true,        
 pdffitwindow=true,     
 pdfstartview={FitH},    
 pdfsubject={Neutrino Oscillations Phenomenology},   
 pdfnewwindow=true,      
 pdfcreator={RevTeX},
 colorlinks=true,       
 linkcolor=red,          
 citecolor=blue,        
 filecolor=black,      
 urlcolor=blue,           
}
  
\newcommand{\ie}{\textit{i.e.}}
\newcommand{\eg}{{\it e.g.}}
\newcommand{\fig}{Fig.}

\newcommand{\Refe}{Ref.}
\newcommand{\Refes}{Refs.}

\newcommand{\dcp}{\delta_{\mathrm{CP}}}

\newcommand{\equ}[1]{Eq.~(\ref{equ:#1})}
\newcommand{\figu}[1]{\fig~\ref{fig:#1}}


\preprint{IP/BBSR/2023-04}

\title{Flavor-dependent long-range neutrino interactions in DUNE \& T2HK: alone they constrain, together they discover}

\author[a,b]{\orcidlink{0000-0002-8363-7693}Masoom Singh,}
\author[c]{\orcidlink{0000-0001-6923-0865}Mauricio Bustamante,}
\author[a,d,e]{\orcidlink{0000-0002-9714-8866}Sanjib Kumar Agarwalla}

\affiliation[a]{Institute of Physics, Sachivalaya Marg, Sainik School Post, Bhubaneswar 751005, India}
\affiliation[b]{Department of Physics, Utkal University, Vani Vihar, Bhubaneswar 751004, India}
\affiliation[c]{Niels Bohr International Academy, Niels Bohr Institute, University of Copenhagen, DK-2100 Copenhagen, Denmark}
\affiliation[d]{Homi Bhabha National Institute, Training School Complex, Anushakti Nagar, Mumbai 400094, India}
\affiliation[e]{Department of Physics \& Wisconsin IceCube Particle Astrophysics Center, University of Wisconsin, Madison, WI 53706, U.S.A}

\emailAdd{masoom@iopb.res.in}
\emailAdd{mbustamante@nbi.ku.dk} 
\emailAdd{sanjib@iopb.res.in } 

\abstract{Discovering new neutrino interactions would represent evidence of physics beyond the Standard Model.  We focus on new flavor-dependent long-range neutrino interactions mediated by ultra-light mediators, with masses below $10^{-10}$~eV, introduced by new lepton-number gauge symmetries $L_e-L_\mu$, $L_e-L_\tau$, and $L_\mu-L_\tau$.  Because the interaction range is ultra-long, nearby and distant matter --- primarily electrons and neutrons --- in the Earth, Moon, Sun, Milky Way, and the local Universe, may source a large matter potential that modifies neutrino oscillation probabilities.  The upcoming Deep Underground Neutrino Experiment (DUNE) and the Tokai-to-Hyper-Kamiokande (T2HK) long-baseline neutrino experiments will provide an opportunity to search for these interactions, thanks to their high event rates and well-characterized neutrino beams.  We forecast their probing power.  Our results reveal novel perspectives.  Alone, DUNE and T2HK may strongly constrain long-range interactions, setting new limits on their coupling strength for mediators lighter than $10^{-18}$~eV.  However, if the new interactions are subdominant, then both DUNE and T2HK, together, will be needed to discover them, since their combination lifts parameter degeneracies that weaken their individual sensitivity.  DUNE and T2HK, especially when combined, provide a valuable opportunity to explore physics beyond the Standard Model.}


\arxivnumber{2305.05184}

\begin{document}
\maketitle
\flushbottom

\section{Introduction}
\label{sec:introduction}

\begin{figure}[t!]
\centering
 \includegraphics[width=0.8\linewidth]{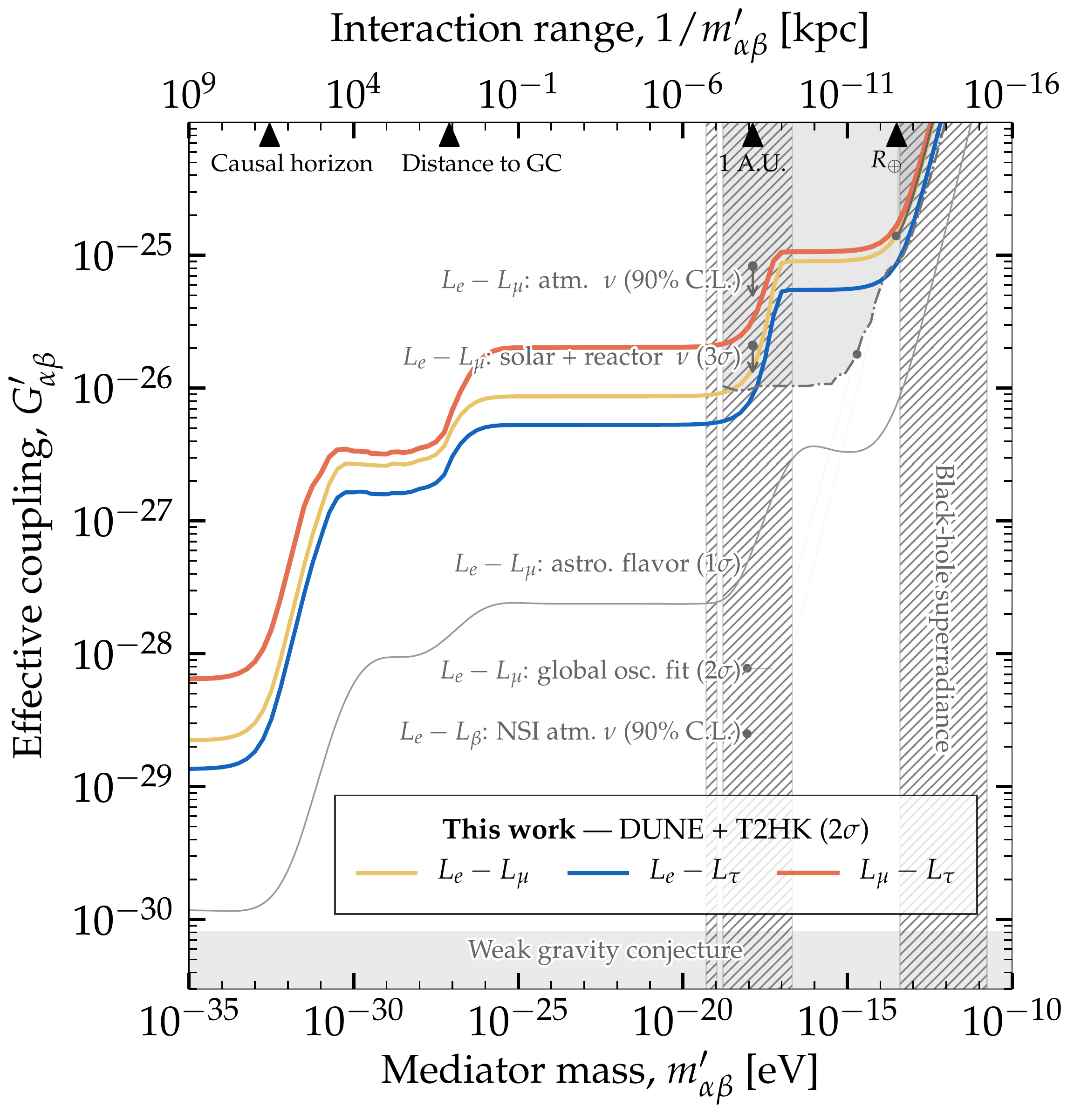}
 \caption{\textbf{\textit{Projected upper limits on the effective coupling, $G_{\alpha\beta}^\prime$ (Eq.~(\ref{equ:Gab})), of the new boson, $Z_{\alpha\beta}^\prime$, with mass $m_{\alpha\beta}^\prime$, that mediates flavor-dependent long-range neutrino interactions, using DUNE, T2HK, and their combination.}}  DUNE runs for 5 years in $\nu$ mode and 5 years in $\bar{\nu}$ mode.  T2HK runs for 2.5 years in $\nu$ mode and 7.5 years in $\bar{\nu}$ mode.  For this plot, we assume that the neutrino mass ordering is normal. Existing limits are from a recent global oscillation fit~\cite{Coloma:2020gfv} (2$\sigma$), atmospheric neutrinos~\cite{Joshipura:2003jh} (90\% C.L.), solar and reactor neutrinos~\cite{Bandyopadhyay:2006uh} (3$\sigma$), and non-standard interactions~\cite{Super-Kamiokande:2011dam, Ohlsson:2012kf, Gonzalez-Garcia:2013usa} (90\% C.L.).  We show the projected sensitivity ($1\sigma$) expected from flavor-composition measurements of high-energy astrophysical neutrinos in IceCube-Gen2~\cite{Bustamante:2018mzu}.  These limits are for the $L_e - L_\mu$ symmetry; see \figu{constraints_g_vs_m} for others.  Indirect limits~\cite{Wise:2018rnb} are from black-hole superradiance (90\% C.L.)~\cite{Baryakhtar:2017ngi}, and the weak gravity conjecture~\cite{Arkani-Hamed:2006emk}, assuming a lightest neutrino mass of $0.01$~eV.  {\it Our projected limits may improve on existing ones, especially for ultra-light mediators of masses below about $10^{-18}$~eV.}  See Section~\ref{sec:results_constraints} for details,  \figu{constraints_g_vs_m} for constraints using DUNE or T2HK separately, and Figs.~\ref{fig:g_vs_m_discovery}--\ref{fig:ranges_th23_V} for discovery plots.
 }
 \label{fig:moneyplot}
\end{figure}
%
Discovering a new fundamental interaction would be striking evidence of physics beyond the Standard Model.  Yet, because new interactions are likely feeble, they are difficult to detect.  And because they may manifest in a variety of ways, they are difficult to search for comprehensively.  So far, there is no evidence for them, despite a long history of searches, though there are stringent limits on their strength~\cite{Lee:1955vk, Okun:1995dn, Williams:1995nq, Dolgov:1999gk, Adelberger:2003zx, Williams:2004qba}.

Starting in the 2030s, the next-generation long-baseline neutrino experiments, Deep Underground Neutrino Experiment (DUNE)~\cite{DUNE:2021cuw} and Tokai-to-Hyper-Kamiokande (T2HK)~\cite{Hyper-Kamiokande:2016srs, Hyper-Kamiokande:2018ofw}, presently under construction, will bring about an opportunity to search for new physics, via neutrinos, more incisively than ever before.  Neutrinos have immense potential to reveal new physics~\cite{Arguelles:2019xgp, Arguelles:2022tki, Berryman:2022hds, Coloma:2022dng, Huber:2022lpm}.  DUNE (\eg, \Refes~\cite{Altmannshofer:2019zhy,Ballett:2019xoj,DeRomeri:2019kic,DUNE:2020fgq,Schwetz:2020xra,Mathur:2021trm,Capozzi:2021nmp,Dev:2021qjj,Chakraborty:2021apc,Ovchynnikov:2022rqj,Chauhan:2022iuh,Denton:2022pxt,Agarwalla:2023wft}) and T2HK (\eg, \Refes~\cite{Farzan:2016wym,Agarwalla:2017nld,Tang:2017khg,Pasquini:2018udd,Ding:2019zhn,Pal:2019pqh,Chakraborty:2020cfu,Hyper-Kamiokande:2022smq}) target this potential via rich physics programs, both within the standard neutrino paradigm and beyond it, that stem from their high expected event rates and well-characterized neutrino beams.  We focus on their capability to look for new neutrino-matter interactions: because, in the Standard Model, neutrinos interact only weakly, the presence of an additional neutrino interaction may be more easily spotted, even if it is feeble.  

We consider flavor-dependent neutrino-matter interactions, originally introduced in \Refes~\cite{He:1990pn,Foot:1990uf, Foot:1990mn, He:1991qd, Foot:1994vd}, and explored and constrained in earlier literature, \eg, in \Refes~\cite{Joshipura:2003jh, Grifols:2003gy, Bandyopadhyay:2006uh, Honda:2007wv, Langacker:2008yv, Heeck:2010pg, Davoudiasl:2011sz, Chatterjee:2015gta, Farzan:2016wym, Wise:2018rnb, Bustamante:2018mzu, Heeck:2018nzc, Khatun:2018lzs, Smirnov:2019cae, Joshipura:2019qxz, KumarPoddar:2019ceq, Coloma:2020gfv, KumarPoddar:2020kdz,Dror:2020fbh, Melas:2023olz,Alonso-Alvarez:2023tii}.  Two reasons motivate our choice.  First, if these interactions are long-range, \ie, if they act across long distances, then large collections of nearby and distant matter may source a sizable matter potential that affects neutrino flavor oscillations appreciably.  Thus, we concentrate on interactions mediated by new, ultra-light mediators, with masses below $10^{-10}$~eV, that subtend ultra-long interaction ranges.  Second, the flavor-dependent interactions we consider are born from gauging, anomaly-free, global symmetries of the Standard Model~\cite{Pontecorvo:1957qd, Pontecorvo:1967fh, Gribov:1968kq, Hisano:1998fj, Cirigliano:2005ck, Altarelli:2010gt}: $L_e-L_\mu$, $L_e-L_\tau$, and $L_\mu-L_\tau$, where $L_e$, $L_\mu$, and $L_\tau$ are the electron, muon, and tau lepton numbers.  This makes them arguably natural and economical extensions of the Standard Model.  Gauging each one introduces a single new neutral vector boson that mediates new neutrino interactions with electrons or neutrons (interactions with other particles are suppressed, as we elaborate on later).

Previous works have explored the sensitivity of existing and future long-baseline neutrino experiments to flavor-dependent long-range interactions.  However, they either fixed the interaction range, typically to be equal to the Sun-Earth distance (see, \eg, \Refe~\cite{Chatterjee:2015gta}), or considered mediator masses only as small as about $10^{-18}$~eV (see, \eg, \Refe~\cite{Coloma:2020gfv}).  We abandon both limitations and explore mediator masses down to $10^{-35}$~eV.  Doing so opens up a largely unexplored regime of ultra-long-range interactions.  As pointed out in \Refe~\cite{Bustamante:2018mzu}, a mediator this light allows for electrons and neutrons in the Earth, Moon, Sun, Milky Way, and the cosmological distribution of matter to affect neutrino oscillations.  To make our forecasts realistic, we base them on detailed simulations of DUNE and T2HK, including their different detection channels, efficiency, backgrounds, and run times.  

Figure~\ref{fig:moneyplot} conveys the novel perspectives revealed by our work.  It shows the first half of our main results, concerning constraints: \textbf{\textit{separately or, as in \figu{moneyplot}, together, DUNE and T2HK may place the strongest constraints on long-range interactions, especially for mediators lighter than $10^{-18}$~eV}}.  (Future sensitivity from flavor measurements of high-energy astrophysical neutrinos in the IceCube-Gen2 neutrino telescope might be comparable~\cite{Bustamante:2018mzu}, but, for now, they are subject to large uncertainties in the neutrino flux, not pictured in \figu{moneyplot}, unlike the constraints from DUNE and T2HK.)  The other half of our main results, not contained in \figu{moneyplot}, concerns discovery.  We find that, separately, DUNE and T2HK will likely be unable to discover subdominant flavor-dependent long-range neutrino interactions, due to degeneracies between their effect on neutrino oscillations and that of the standard mixing parameters.  Yet, \textbf{\textit{together, their complementary capabilities may lift degeneracies and enable the discovery of the new interactions; see \figu{g_vs_m_discovery}.}}  Below, we elaborate on these perspectives.

This paper is organized as follows.  Section~\ref{sec:lri} introduces lepton-number gauge symmetries, long-range interactions, and their effect on neutrino oscillations.  Section~\ref{sec:experiments} overviews DUNE and T2HK, and shows oscillation probabilities and event rates in them.  Section~\ref{sec:results} shows projected constraints and discovery prospects.  Section~\ref{sec:conclusions} summarizes and concludes. 
%
\section{Flavor-dependent long-range neutrino  interactions}
\label{sec:lri}
\subsection{Gauged lepton-number symmetries}
\label{sec:lri_symmetries}
\begin{figure}
\includegraphics[width=\linewidth]{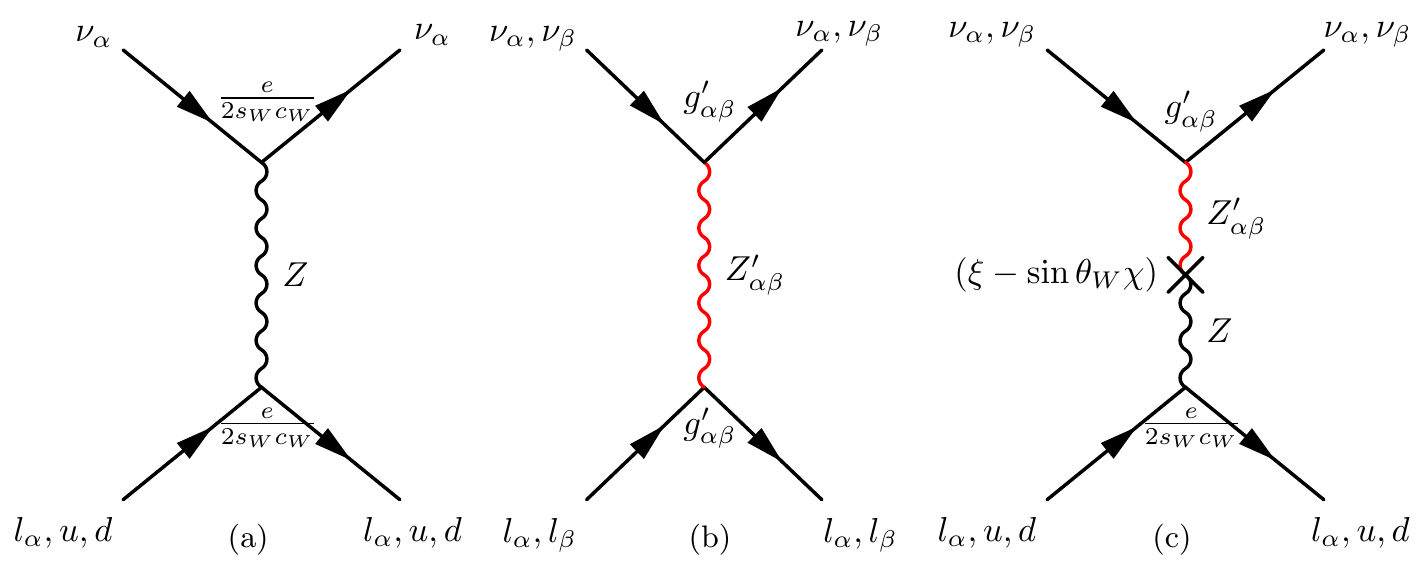}
\caption{\textbf{\textit{Feynman diagrams for neutrino-matter interactions.}}  Each diagram corresponds to a term in the Lagrangian, \equ{lagrangian_total}: (a) SM contribution mediated by the $Z$ boson, (b) new contribution from the gauge symmetry $U(1)_{L_\alpha-L_\beta}$, mediated by the new boson $Z_{\alpha\beta}^\prime$, and (c) mixing between $Z$ and $Z_{\alpha\beta}^\prime$.  In our analysis, (a) is significant only for neutrinos inside the Earth. For $L_{e} - L_\beta$ symmetries, (b) is the only additional contribution sourced by electrons.  For $L_\mu - L_\tau$, (c) is instead the only additional contribution sourced by neutrons.  See Section~\ref{sec:lri_symmetries} for details.}
\label{fig:feynman}
\end{figure}
%
In the Standard Model (SM), the baryon number and the lepton numbers, $L_e$, $L_\mu$, and $L_\tau$, are accidental $U(1)$ global symmetries.  Linear combinations of the lepton-number symmetries can be gauged anomaly-free, \ie, without introducing a new fermion or right-handed neutrino (although not simultaneously)~\cite{Foot:1990mn, Foot:1990uf, He:1991qd, Foot:1994vd}.  To showcase the capabilities of DUNE and T2HK, we explore three such new $U(1)$ gauge symmetries, generated by $L_e-L_\mu$, $L_e-L_\tau$, and $L_\mu-L_\tau$, that introduce new flavor-dependent neutrino-matter interactions; later, we show how they affect neutrino oscillations.  (Other combinations of baryon and lepton numbers can also be gauged anomaly-free; see \Refe~\cite{Coloma:2020gfv}.) 

Figure~\ref{fig:feynman} shows the Feynman diagrams for neutrino-matter interaction that we consider.  For a particular lepton-number symmetry, the corresponding effective Lagrangian is
\begin{equation}
 \label{equ:lagrangian_total} 
 \mathscr{L}_{\rm eff}
 =
 \mathscr{L}_{\rm SM}
 +
 \mathscr{L}_{Z^\prime}
 +
 \mathscr{L}_{\text{mix}} \;.
\end{equation}
The first term describes the SM contribution, mediated by the $Z$ boson, \ie,
\begin{equation}
 \label{equ:lagrangian_sm}
 \mathscr{L}_{\text{SM}}
 =
 \frac{e}{\sin \theta_{W} \cos \theta_{W}}Z_\mu 
 \left[-\frac{1}{2}\bar{l}_{\alpha}\gamma^\mu P_{L} l_{\alpha}+\frac{1}{2}\bar{\nu}_{\alpha}\gamma^\mu P_{L} \nu_{\alpha}+\frac{1}{2}\bar{u}\gamma^\mu P_{L} u-\frac{1}{2}\bar{d}\gamma^\mu P_{L} d
 \right] \;,
\end{equation}
where $e/(\sin\theta_W\cos\theta_W) = 0.723$, $e$ is the unit charge, $\theta_W$ is the Weinberg angle, $\nu_\alpha$ and $l_\alpha$ are a neutrino and charged lepton of flavor $\alpha = e, \mu, \tau$, $P_{L}$ is the left-handed projection operator, and $u$ and $d$ are up and down quarks.  Because the $Z$ boson is heavy, the interaction that it mediates is short-range; in our work, it matters only inside the Earth.  (Equation~(\ref{equ:lagrangian_sm}), and also \equ{lag_mix} below, assumes that matter is electrically neutral, \ie, that it has equal abundance of electrons and protons~\cite{Heeck:2010pg}, which is also what we assume later when computing the new matter potential; see Section~\ref{sec:lri_potential}.)

The second term in \equ{lagrangian_total} describes the interaction between $\nu_\alpha$ and $l_\alpha$ mediated by the new $Z_{\alpha\beta}^\prime$ boson~\cite{He:1990pn, He:1991qd, Heeck:2010pg}, \ie, for the $L_\alpha-L_\beta$ symmetry,
\begin{equation}
 \label{equ:lagrangian_zprime}
 \mathscr{L}_{Z^\prime}
 =
 g_{\alpha \beta}^\prime Z^\prime_{\sigma}(\bar{l}_{\alpha}\gamma^{\sigma}l_{\alpha}-\bar{l}_{\beta}\gamma^{\sigma}l_{\beta}+ \bar{\nu}_{\alpha}\gamma^{\sigma}P_{L}\nu_{\alpha}-\bar{\nu}_{\beta}\gamma^{\sigma}P_{L}\nu_{\beta}) \;,
\end{equation}
where $g_{\alpha\beta}^\prime$ is a dimensionless coupling constant.  Due to the dearth of naturally occurring muons and tauons with which neutrinos can interact, we neglect this contribution under $L_\mu-L_\tau$ and consider it only under $L_e-L_\mu$ and $L_e-L_\tau$, for which the interaction is sourced by comparatively abundant electrons.

The final term in \equ{lagrangian_total} describes the mixing between $Z$ and $Z_{\alpha\beta}^\prime$~\cite{Babu:1997st, Heeck:2010pg, Joshipura:2019qxz}, which can arise directly or by radiative mixing~\cite{Holdom:1985ag,Tomalak:2020zfh}.  In the physical basis, this term is~\cite{Babu:1997st} $\mathscr{L}_{ZZ^\prime} \supset
 (\xi-\sin\theta_W\chi) Z'_{\mu}Z^{\mu}$,
where $\chi$ is the kinetic mixing angle between the two bosons and $\xi$ is the rotation angle between gauge eigenstates and physical states.  This introduces a four-fermion interaction between neutrinos and charged leptons, protons, and neutrons via $Z$--$Z_{\alpha\beta}^\prime$ mixing, \ie,
\begin{equation}
 \label{equ:lag_mix}
 \mathcal{L}_{\rm mix}
 =
 -g_{\alpha\beta}^\prime
 (\xi-\sin\theta_W\chi)\frac{e}{\sin\theta_W \cos\theta_W}
 J'_\rho J_3^\rho \;,
\end{equation}
where $J^\prime_\rho = \bar{\nu}_\alpha \gamma_\rho P_L\nu_\alpha-\bar{\nu}_\beta \gamma_\rho P_L\nu_\beta$ and $J_3^\rho = -\frac{1}{2}\bar{e}\gamma^\rho P_L e+\frac{1}{2}\bar{u}\gamma^\rho P_L u-\frac{1}{2}\bar{d}\gamma^\rho P_L d$.
However, the contribution of electrons is nullified by that of protons, leaving only neutrons to source the new interaction via mixing.  The term $(\xi - \sin\theta_{W}\chi)$ effectively describes the strength of the $Z$--$Z_{\alpha\beta}^\prime$ mixing.  Its value is unknown, but there are upper limits on it~\cite{Schlamminger:2007ht, Adelberger:2009zz, Heeck:2010pg}. We do not consider its value independently, but together with $g_{\alpha\beta}^{\prime}$\,, as an effective coupling (more on this below). In order to showcase the effect of mixing, we include $\mathscr{L}_{\rm mix}$ only under $L_\mu-L_\tau$.

In summary, under $L_e - L_\mu$ and $L_e - L_\tau$, the new interactions are described by $\mathscr{L}_{Z^\prime}$, and are sourced by electrons only, whereas under $L_\mu-L_\tau$, the new interactions are described by $\mathscr{L}_{\rm mix}$, and are sourced by neutrons only.  In all cases, in addition, standard neutrino-electron interactions, described by $\mathscr{L}_{\rm SM}$, are active only inside the Earth.
\subsection{Long-range matter potential}
\label{sec:lri_potential}
The above interactions induce flavor-dependent Yukawa potentials, sourced by electrons and neutrons, that affect the mixing of neutrinos~\cite{He:1990pn, Foot:1990mn, He:1991qd, Foot:1994vd, Bustamante:2018mzu, Wise:2018rnb}.  Under $L_e-L_\beta$ ($\beta = \mu, \tau$), a neutrino located at a distance $d$ from a collection of $N_e$ electrons experiences a potential
\begin{equation}
 V_{e\beta}
 =
 G_{\alpha\beta}^{\prime 2}
 \frac{N_e}{4\pi d}
 e^{-m'_{e\beta}d} \;,
 \label{equ:potential_e_beta}
\end{equation}
where $m_{e\beta}^\prime$ is the mass of the mediating $Z_{e\beta}^\prime$ boson.  Under $L_\mu-L_\tau$\,, a neutrino located at a distance $d$ from a collection of $N_n$ neutrons experiences a potential
\begin{equation}
 V_{\mu\tau}
 =
 G_{\alpha\beta}^{\prime 2}
  \frac{e}{\sin\theta_W\cos\theta_W}
 \frac{N_n}{4 \pi d}
 e^{-m'_{\mu\tau}d} \;,
 \label{equ:potential_mu_tau}
\end{equation}
where $m_{\mu\tau}^\prime$ is the mass of the mediating $Z_{\mu\tau}^\prime$ boson. In Eqs.~(\ref{equ:potential_e_beta}) and~(\ref{equ:potential_mu_tau}), the effective coupling strength is
\begin{equation}
 G_{\alpha \beta}^\prime
 =
 \left\{
  \begin{array}{lll}
   g^{\prime }_{e \mu} & , & ~{\rm for}~\alpha, \beta = e, \mu \\
   g^{\prime }_{e \tau} & , & ~{\rm for}~\alpha, \beta = e, \tau \\
   \sqrt{g^{\prime}_{\mu \tau} (\xi-\sin \theta_W \chi)} & , & ~{\rm for}~\alpha, \beta = \mu, \tau \\
  \end{array}
 \right. \;.
 \label{equ:Gab}
\end{equation}
At distances longer than the interaction range of $1/m_{\alpha\beta}^\prime$, the potential is suppressed due to the mediator mass.  Like \Refe~\cite{Bustamante:2018mzu}, we explore light mediators with $m_{\alpha\beta}^\prime = 10^{-35}$--$10^{-10}$~eV, corresponding to interaction ranges from $10^3$~Gpc --- larger than the observable Universe --- to hundreds of meters; see \figu{moneyplot}.

We adopt the methods introduced in \Refe~\cite{Bustamante:2018mzu} to compute the total potential sourced by nearby and faraway electrons and neutrons in the Earth ($\oplus$), Moon ($\leftmoon$), Sun ($\astrosun$), Milky Way (MW), and by the cosmological distribution of matter (cos) in the local Universe, \ie,
\begin{equation}
 \label{equ:pot_total}
 V_{\alpha \beta}
 =
 V_{\alpha \beta}^\oplus + V_{\alpha \beta}^{\leftmoon} + V_{\alpha \beta}^{\astrosun} + V_{\alpha \beta}^{\rm MW} +  V_{\alpha \beta}^{\rm cos} \;.
\end{equation} 
The specific value of $m_{\alpha\beta}^\prime$ determines the relative size of the contributions of the above sources to the total potential.  We do not compute the changing potential along the underground trajectories of the neutrinos from source to detector inside the Earth; see \Refe~\cite{Coloma:2020gfv} for such treatment. Instead, like \Refe~\cite{Bustamante:2018mzu}, we compute the average potential experienced by the neutrinos at their point of detection.  This approximation is especially valid for mediators lighter than about $10^{-14}$~eV, for which the interaction range is longer than the radius of the Earth (see \figu{moneyplot}), and so all of the electrons and neutrons on Earth contribute to the potential experienced by a neutrino regardless of its position along its trajectory.  Below $10^{-14}$~eV is also where we place novel projected limits.

We assume that the matter that sources the potential is electrically neutral, so that the number of electrons and protons is the same, and isoscalar, so that the number of electrons and neutrons is the same, except for the Sun~\cite{Heeck:2010pg} and for the cosmological distribution of matter~\cite{Hogg:1999ad, Steigman:2007xt, Planck:2015fie}.  We treat the Moon ($N_{e,\leftmoon} = N_{n,\leftmoon}\sim 5 \cdot 10^{49}$) and the Sun ($N_{e,\astrosun} \sim 10^{57}$, $N_{n,\astrosun} = N_{e,\astrosun}/4$) as point sources of electrons and neutrons, and the Earth ($N_{e, \oplus} \approx N_{n, \oplus} \sim 4 \times 10^{51}$), the Milky Way ($N_{e, {\rm MW}} \approx N_{n, {\rm MW}} \sim 10^{67}$), and the cosmological matter ($N_{e,\mathrm{cos}} \sim 10^{79}$, $N_{n,\mathrm{cos}} \sim 10^{78}$) as continuous distributions. We defer to \Refe~\cite{Bustamante:2018mzu} for a detailed calculation of \equ{pot_total}, but adopt two differences introduced by \Refe~\cite{Agarwalla:2023sng}.  First, unlike \Refe~\cite{Bustamante:2018mzu}, which studied extragalactic neutrinos and so averaged the contribution of cosmological matter over redshift, here we consider $V_{\alpha\beta}^{\rm cos}$ to be only the contribution from the local Universe, \ie, we evaluate Eq.~(A8) in \Refe~\cite{Bustamante:2018mzu} at redshift $z = 0$.  Second, unlike \Refe~\cite{Bustamante:2018mzu}, which only computed the potential sourced by electrons under $L_e-L_\mu$ and $L_e-L_\tau$, here we compute also the potential sourced by neutrons under $L_\mu-L_\tau$.
%
\subsection{Neutrino oscillation probabilities under long-range interactions}
\label{sec:lri_osc_prob}
We consider mixing between the three active neutrinos, $\nu_e$, $\nu_\mu$, and $\nu_\tau$.  Under the $L_\alpha-L_\beta$ symmetry, the Hamiltonian that drives neutrino propagation, in the flavor basis, is
\begin{equation}
 \label{equ:hamiltonian_tot}
 \mathbf{H}
 =
 \mathbf{H}_{\rm vac}
 +
 \mathbf{V}_{\rm mat}
 +
 \mathbf{V}_{\alpha\beta} \;.
\end{equation}
The first two terms on the right-hand side induce standard oscillations, including SM matter effects; the third one, oscillations due to the new interactions.  

In vacuum, oscillations are driven by
\begin{equation}
 \mathbf{H}_{\rm vac}
 =
 \frac{1}{2 E}
 \mathbf{U}~
 {\rm diag}(0, \Delta m^2_{21}, \Delta m^2_{31})
 ~\mathbf{U}^{\dagger} \;,
\end{equation}
where $E$ is the neutrino energy, $\Delta m^2_{ij} \equiv m^2_i-m^2_j$ are the mass-squared splittings between neutrino mass eigenstates, and $\mathbf{U}$ is the Pontecorvo-Maki-Nakagawa-Sakata (PMNS) mixing matrix, parametrized~\cite{ParticleDataGroup:2022pth} in terms of the  mixing angles $\theta_{12}$, $\theta_{23}$, and $\theta_{13}$, and the CP-violating phase, $\dcp$.  In the main text, we show results assuming normal neutrino mass ordering (NMO), where $m_1 < m_2 < m_3$; Table~\ref{tab:mix_param_benchmark} shows the values of the mixing parameters that we use, taken from \Refe~\cite{Capozzi:2021fjo}.  Appendix~\ref{app:imo} contains results obtained under the inverted mass ordering (IMO), where $m_3 < m_2 < m_1$.

In \equ{hamiltonian_tot}, the contribution of SM coherent forward scattering on electrons, mediated by the $W$ boson, is
\begin{equation}
 \mathbf{V}_{\rm mat}
 =
 {\rm diag}(V_{\rm CC}, 0, 0) \;,
\end{equation}
where $V_{\rm CC} = \sqrt{2} G_F n_e \simeq 7.6 Y_e [ \rho / (10^{14}~{\rm g}~{\rm cm}^{-3})]$~eV is the charged-current neutrino-electron interaction potential, $G_F$ is the Fermi constant, $n_e$ is the electron number density, $Y_e \equiv n_e / (n_p + n_n)$ is the electron fraction, \ie, its abundance relative to that of protons and neutrons, $n_p$ and $n_n$, and $\rho$ is the matter density.  In our work, this contribution is relevant only inside Earth, where matter densities are high.  We take $\rho$ to be the average density of underground matter along the trajectory from source to detector, calculated using the Preliminary Reference Earth Model~\cite{Dziewonski:1981xy}: $2.848$~g~cm$^{-3}$ for DUNE and 2.8~g~cm$^{-3}$ for T2HK.  The potential above is for neutrinos; for antineutrinos, it flips sign, \ie, $\mathbf{V}_{\rm mat} \to -\mathbf{V}_{\rm mat}$.

Finally, in \equ{hamiltonian_tot} the contribution from the new matter interaction is
\begin{equation}
 \label{equ:pot_lri_matrix}
 \mathbf{V}_{\alpha\beta}
 =
 \left\{
  \begin{array}{ll}
   {\rm diag}(V_{e\mu}, -V_{e\mu}, 0), & {\rm for}~ \alpha, \beta = e, \mu \\
   {\rm diag}(V_{e\tau}, 0, -V_{e\tau}), & {\rm for}~ \alpha, \beta = e, \tau \\
   {\rm diag}(0, V_{\mu\tau}, -V_{\mu\tau}), & {\rm for}~ \alpha, \beta = \mu, \tau \\   
  \end{array}
 \right. \;,
\end{equation}
where the potential, $V_{\alpha\beta}$, \equ{pot_total}, depends on the mediator mass, $m_{\alpha\beta}^\prime$, and coupling, $G_{\alpha\beta}^\prime$.  The potential above is for neutrinos; for antineutrinos, it flips sign, \ie, $\mathbf{V}_{\alpha\beta} \to -\mathbf{V}_{\alpha\beta}$.

The $\nu_\alpha \to \nu_\beta$ transition probability associated to the Hamiltonian, \equ{hamiltonian_tot}, is
\begin{equation}
 \label{equ:osc_prob}
 P_{\nu_{\alpha}\rightarrow \nu_{\beta}}
 =
 \Bigg| 
 \sum_{i=1}^{3} 
 U^\prime_{\alpha i}
 \exp \left( \dfrac{\Delta \tilde{m}^2_{i1}L}{2E} \right)
 U^{\prime \ast}_{\beta i}
 \Bigg|^{2} \;,
\end{equation}
where $L$ is the distance traveled by the neutrino from production to detection, $\Delta \tilde{m}^2_{ij} \equiv \tilde{m}_i^2 - \tilde{m}_j^2$, with $\tilde{m}_i^2/2E$ the eigenvalues of the Hamiltonian, modified from those of $\mathbf{H}_{\rm vac}$ by matter effects, and $\mathbf{U}^\prime$ is the unitary matrix that diagonalizes the Hamiltonian.   We parametrize $\mathbf{U}^\prime$ with the same shape as the PMNS matrix, but evaluated at mixing parameters $\theta_{12}^m$, $\theta_{23}^m$, $\theta_{13}^m$, and $\dcp^m$ modified by matter effects.  In our work, we compute the oscillation probability, \equ{osc_prob}, exactly and numerically to arbitrary precision; see \Refes~\cite{Barger:1980tf, Zaglauer:1988gz, Ohlsson:1999um, Akhmedov:2004ny, Agarwalla:2013tza, Chatterjee:2015gta, Agarwalla:2015cta, Khatun:2018lzs, Agarwalla:2021zfr} for approximate analytical solutions.

For the new matter interactions to affect the oscillation probability, the new matter potential must be at least comparable to the standard contributions in \equ{hamiltonian_tot}, \ie, in vacuum, $V_{\alpha \beta} \gtrsim (\Delta m_{31}^2/2E)$ [inside the Earth, this is instead $V_{\alpha \beta} \gtrsim \max \left( \Delta m_{31}^2/2E, V_{\rm CC} \right)$].  In DUNE and T2HK, where the first oscillation maxima occur at 2.6~GeV and 0.6~GeV, respectively, this implies that they become important for $V_{\alpha\beta} \gtrsim 10^{-13}$~eV.  This sets the scale of the potential to which our analysis is sensitive.  Later, in Section~\ref{sec:experiments_probabilities}, we show how the new interactions affect the probabilities in DUNE and T2HK.
%
\subsection{Existing limits}
\label{sec:prev_limits}
%
Figure~\ref{fig:moneyplot} (also \figu{constraints_g_vs_m}) shows existing limits on flavor-dependent long-range neutrino interactions.  Below, we summarize them.  We focus on light mediators; the complementary case for heavy mediators was first studied in \Refes~\cite{Foot:1990mn, He:1990pn, He:1991qd, Foot:1994vd, Dutta:1994dx}.

Pioneering studies in \Refes~\cite{Joshipura:2003jh} and \cite{Bandyopadhyay:2006uh} identified the potential of neutrino oscillations to test new long-range interactions, possibly more stringently than gravitational probes.  They focused on interactions with a range equal to the Earth-Sun distance [though ignoring the Yukawa suppression in \equ{potential_e_beta}] and sourced by solar electrons.  Reference~\cite{Joshipura:2003jh} used Super-Kamiokande atmospheric neutrino data to find $g_{e\mu}^\prime < 8.32 \times 10^{-26}$ and $g_{e\tau}^\prime < 8.97 \times 10^{-26}$, at 90\%~confidence level (C.L.)  Reference~\cite{Bandyopadhyay:2006uh} used solar and reactor neutrino data from KamLAND to find $g_{e\mu}^\prime < 2.06 \times 10^{-26}$ and $g_{e\tau}^\prime < 1.77 \times 10^{-26}$, at 3$\sigma$, assuming $\theta_{13} = 0^{\circ}$.

Reference~\cite{Heeck:2010pg} studied the effect of the $L_\mu-L_\tau$ symmetry via kinetic mixing (see Section~\ref{sec:lri_symmetries}) on $\nu_\mu$ in the long-baseline experiment MINOS.  By comparing the potential $V_{\mu\tau}$ sourced by a neutron in the Sun to the fifth-force gravitational potential sourced by it, and applying upper limits on the strength of the latter from torsion-balance experiments~\cite{Schlamminger:2007ht, Adelberger:2009zz}, \Refe~\cite{Heeck:2010pg} set an upper limit on the mixing strength of $(\xi - \sin\theta_W\chi) < 5 \times 10^{-24}$ at 95\%~C.L.~for a long-range interaction with range equal to the Earth-Sun distance.  (This is the limit that we saturate  when computing the $V_{\mu\tau}$ potential, \equ{potential_mu_tau}; see Section~\ref{sec:lri_symmetries}.)  This translates into an upper limit of $g_{\mu\tau}^\prime \leq 2.51 \times 10^{-26}$.  For an interaction with a range of the size of the Earth, the upper limit degrades to $g_{\mu\tau}^\prime \leq 10^{-24}$.  

Reference~\cite{Wise:2018rnb} showed that upper limits on the coefficients that parametrize the strength of non-standard neutrino interactions (NSI) can be translated into upper limits on the coupling strength of flavor-dependent long-range interactions.  Figure~\ref{fig:moneyplot} shows the resulting limits, based on the NSI limits from \Refes~\cite{Super-Kamiokande:2011dam, Ohlsson:2012kf, Gonzalez-Garcia:2013usa}.

Recently, \Refe~\cite{Coloma:2020gfv} performed a global oscillation analysis of new $U(1)$ symmetries, including $L_e-L_\mu$ and $L_e-L_\tau$, by using the same experimental data sets used in NuFIT~5.0~\cite{Esteban:2020cvm, NuFIT}. Unlike our analysis, \Refe~\cite{Coloma:2020gfv} computed the changing long-range matter potential due to underground matter in the Earth along the trajectory of the neutrinos.  Their procedure is more detailed than ours for mediators lighter than $10^{-14}$~eV.  However, they explore masses only as low as $10^{-18}$~eV, \ie, an interaction range of 1~A.U.

Reference~\cite{Bustamante:2018mzu} first showed that the flavor composition of TeV--PeV astrophysical neutrinos, \ie, the relative number of $\nu_e$, $\nu_\mu$, and $\nu_\tau$, can be used to probe long-range interactions under $L_e-L_\mu$ and $L_e-L_\tau$ sourced by the same collections of nearby and distant electrons that we consider here.  Reference~\cite{Agarwalla:2023sng} refined the statistical methods and included also $L_\mu-L_\tau$.  The main effect is that, if the potential sourced by electrons or neutrons were to be dominant, oscillations would turn off, and the flavor composition emitted by the astrophysical sources and received at Earth would be the same; see also \Refe~\cite{Farzan:2018pnk}.  Figure~\ref{fig:moneyplot} shows the projected upper limits obtained in \Refe~\cite{Bustamante:2018mzu} based on estimates of flavor-composition measurements in the envisioned IceCube-Gen2 neutrino telescope~\cite{IceCube-Gen2:2020qha}.
  
Finally, following \Refe~\cite{Wise:2018rnb}, \figu{moneyplot} includes two indirect limits.  First, \Refe~\cite{Baryakhtar:2017ngi} excluded three mediator mass windows (``Black-hole superradiance'') by considering the superradiant growth rate of a gravitationally bound accumulation of light vector bosons around selected stellar-mass and supermassive black holes.  Second, \Refe~\cite{Arkani-Hamed:2006emk} placed a tentative lower limit on the coupling (``Weak gravity conjecture'') by studying low-energy effective theories that contain gravity and $U(1)$ gauge fields where at least one particle charged under $U(1)$ is essential for gravity to be the weakest force.  

In \figu{moneyplot}, we show existing limits as they were published in their original references.  Hence, they do not extend to mediators lighter than $10^{-14}$--$10^{-20}$~eV, depending on the limit (except for the proof-of-principle sensitivity based on projected IceCube-Gen2 measurements of the flavor composition~\cite{Bustamante:2018mzu}).  These limits could be recomputed and extended to span lighter mediators, using the same long-range matter potential that we have used, \equ{pot_total}, though doing so lies beyond the scope of this work.

%
\section{Long-range interactions in DUNE and T2HK}
\label{sec:experiments}
%
\subsection{Overview of the experiments}
\label{sec:experiments_overview}
Long-baseline neutrino experiments are powerful probes of neutrino oscillations~\cite{Feldman:2012jdx, Agarwalla:2014fva, Diwan:2016gmz, Giganti:2017fhf}.  Owing to baselines of hundreds of kilometers and well-characterized GeV-scale neutrino beams, they can probe matter effects in oscillations, CP violation, neutrino mass ordering, and a large number of possible new neutrino physics~\cite{Mena:2005ek, Feldman:2012jdx, Agarwalla:2014fva, Diwan:2016gmz, Farzan:2017xzy, Giganti:2017fhf}. Today, long-baseline experiments T2K~\cite{T2K:2023smv} and NO$\nu$A~\cite{NOvA:2021nfi} contribute high-precision data to global oscillation fits~\cite{Esteban:2020cvm, Capozzi:2021fjo}.  In the coming decade, next-generation experiments DUNE~\cite{DUNE:2015lol, DUNE:2020lwj, DUNE:2020ypp, DUNE:2020jqi, DUNE:2021cuw, DUNE:2021mtg}, T2HK~\cite{Hyper-Kamiokande:2016srs, Hyper-Kamiokande:2018ofw}, and the European Spallation Source neutrino Super Beam (ESS$\nu$SB)~\cite{Blennow:2019bvl, ESSnuSB:2021azq}, currently under construction, will take this further~\cite{Mena:2005ek, Feldman:2012jdx, Agarwalla:2014fva, Diwan:2016gmz, Farzan:2017xzy, Giganti:2017fhf}.  
%
\begin{table}[t!]
 \centering
 \begin{tabular}{|c|c|c|c|c|c|c|}
  \hline 
  \multirow{2}{*}{} & \multicolumn{6}{c|}{Standard mixing parameters (NMO)} \\
   & $\sin^2 \theta_{12}$ & $\sin^2\theta_{23}$ & $\sin^2 \theta_{13}$ &
  $\frac{\Delta m^2_{31}}{10^{-3}\,\text{eV}^2}$  & $\frac{\Delta m^2_{21}}{10^{-5}\,\text{eV}^2}$  & $\delta_{\rm CP}\, (^\circ)$\\[0.8ex]
  \hline
  Benchmark & 0.303 & 0.455 & 0.0223 & 2.522 & 7.36 & 223  \\
  Status in fits & Fixed & Minimized & Fixed & Minimized & Fixed & Minimized \\
  Range & -- & [0.4, 0.6] & -- & [2.438, 2.602] & -- & [139, 355]\\
  \hline 
 \end{tabular}
 \caption{\textbf{\textit{Values of the standard mixing parameters used in our analysis.}}  We assume normal neutrino mass ordering (NMO) in the main text.  The benchmark values are the best-fit values from \Refe~\cite{Capozzi:2021fjo}.  For each parameter over which we minimize our test statistic (see Section~\ref{sec:results_stat_methods}), the minimum is searched for within the range shown, which is the $3\sigma$ allowed range from \Refe~\cite{Capozzi:2021fjo}.  We assume no correlation between the parameters. Table~\ref{tab:mix_param_benchmark_imo} shows the parameter ranges that we use in Appendix~\ref{app:imo} to obtain results under the inverted mass ordering (IMO) instead.}
 \label{tab:mix_param_benchmark}
\end{table} 
%

In our forecasts, we focus on DUNE and T2HK.  Below, we overview their features.  For each one, we compute appearance and disappearance oscillation probabilities and event rates in neutrino ($\nu$) and antineutrino ($\bar{\nu}$) beam modes:
\begin{description}
 \item 
  [{\bf Appearance, $\nu$ mode:}] This is sensitive mainly to $\nu_\mu \to \nu_e$ transitions.  The beam works in neutrino mode, and the detector targets $\nu_e$-initiated events.
 \item 
  [{\bf Appearance, $\bar{\nu}$ mode:}] This is sensitive mainly to $\bar{\nu}_\mu \to \bar{\nu}_e$ transitions.  The beam works in antineutrino mode, and the detector targets $\bar{\nu}_e$-initiated events.
 \item 
  [{\bf Disappearance, $\nu$ mode:}] This is sensitive mainly to $\nu_\mu \to \nu_\mu$ survival.  The beam works in neutrino mode, and the detector targets $\nu_\mu$-initiated events.
 \item
  [{\bf Disappearance, $\bar{\nu}$ mode:}] This is sensitive mainly to $\bar{\nu}_\mu \to \bar{\nu}_\mu$ survival.  The beam works in antineutrino mode, and the detector targets $\bar{\nu}_\mu$-initiated events.
\end{description}
DUNE will also detect $\nu_\tau$ with energies larger than 3.4~GeV via their charged-current interactions, which allows for interesting physics opportunities~\cite{Conrad:2010mh, DeGouvea:2019kea, Ghoshal:2019pab, Machado:2020yxl, Agarwalla:2021owd}.  However, in our analysis, we focus on $\nu_e$ appearance only and treat $\nu_\tau$ appearance as background; see below.

Table~\ref{tab:mix_param_benchmark} shows the values and allowed ranges of the mixing parameters that we use in our analysis, taken from the global oscillation fit of \Refe~\cite{Capozzi:2021fjo}.  In the main text, we show results assuming that the true neutrino mass ordering is normal, since there is currently weak preference for it~\cite{deSalas:2020pgw, Esteban:2020cvm, NuFIT, Capozzi:2021fjo}.  However, as part of our statistical analysis in Section~\ref{sec:results}, we report sensitivity after minimizing over the mass ordering.  Appendix~\ref{app:imo} contains results assuming instead that the true mass ordering is inverted.  
\subsubsection{DUNE}
\label{sec:experiments_overview_dune}
DUNE will consist of a near detector, about 600~m downstream of the neutrino production point on the Fermilab site, and a far detector, 1285~km away and about 1.5~km underground, in the Sanford Underground Research Facility in South Dakota~\cite{DUNE:2018tke}.  The near detector will monitor and characterize the neutrino beam (though it has physics capabilities itself, too~\cite{DUNE:2021tad, DUNE:2022aul}).  We focus on the far detector since it offers prime sensitivity to neutrino oscillations.  It is a state-of-the-art liquid-argon time projection chamber with a net volume of 40~kton; to generate our results, we consider single-phase detection only~\cite{DUNE:2020jqi}.  Neutrino detection is via charged-current neutrino-argon interaction.  Detector deployment will be phased~\cite{DUNE:2021mtg}, but in our simulations we consider only the final, total detector volume.  

DUNE will use the Long Baseline Neutrino Facility (LBNF) neutrino beam produced at Fermilab.  There, the Main Injector of the LBNF fires a 1.2-MW beam of protons of 120~GeV onto a graphite target, producing charged mesons that decay in flight to neutrinos. The resulting neutrino flux is wide-band, ranges from a few hundreds of MeV to a few tens of GeV, and is expected to peak at 2.5~GeV, with most neutrinos in the 1--5~GeV range.  By changing the polarity of the focusing horns~\cite{DUNE:2018hrq, DUNE:2018mlo}, the experiment can run in neutrino or antineutrino mode.  Following the DUNE Technical Design Report~\cite{DUNE:2020jqi}, we adopt a run time of 5 years in neutrino mode and 5 years in antineutrino mode.  This amounts to $1.1 \times 10^{21}$ protons-on-target per year and a net exposure of 480 kton~MW~year.  
To produce our results, we use the DUNE simulation configuration from \Refe~\cite{DUNE:2021cuw}.  
%
\begin{figure}[htb!]
 \centering
 \includegraphics[width=\linewidth]{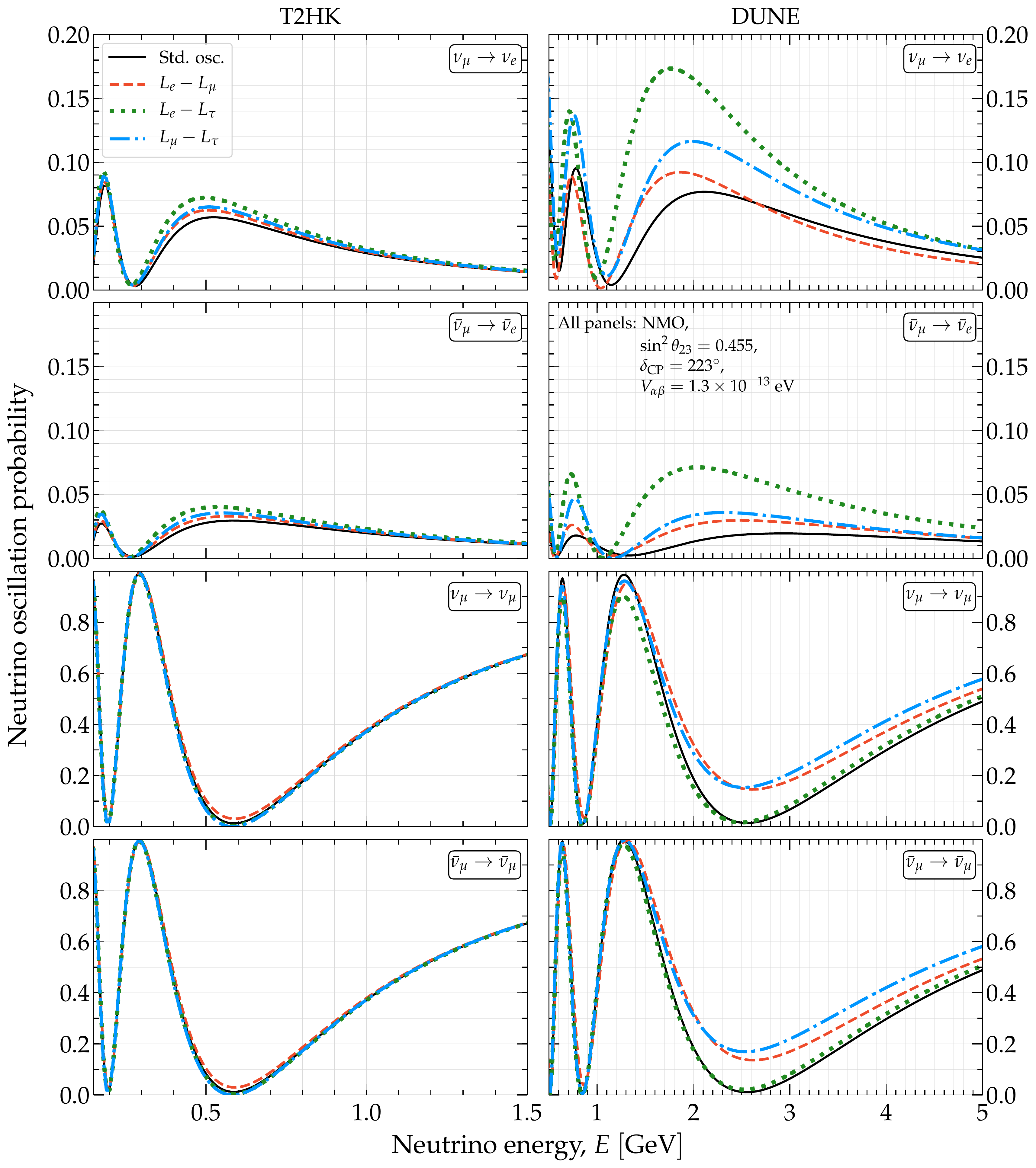}
 \caption{\textbf{\textit{Neutrino oscillation probabilities for T2HK (left column) and DUNE (right column) under flavor-dependent long-range neutrino interactions.}}  The interactions are induced by the lepton-number symmetry $L_e-L_\mu$, $L_e-L_\tau$, or $L_\mu-L_\tau$.  For this figure, we fix the long-range potential to $V_{\alpha\beta} = 1.3 \times 10^{-13}$~eV as illustration, and the standard mixing parameters to their benchmark values from Table~\ref{tab:mix_param_benchmark}.  See Section~\ref{sec:experiments_probabilities} for details.}
 \label{fig:probabilities}
\end{figure}
\subsubsection{T2HK}
T2HK will consist of near detectors, about 280~m downstream from the neutrino production point at the Japan Proton Accelerator Research Complex (JPARC), and a far detector, 295~km away and about 1.7~km underground, in the Tochibora mines of Japan, 8~km from Super-Kamiokande~\cite{Hyper-Kamiokande:2022smq}.  Like in DUNE, the near detectors will monitor and characterize the neutrino beam, and we focus on the far detector.  It will be a tank filled with purified water, with a net volume of 187~kton, whose internal wall is lined with photomultipliers (PMTs).  Neutrino detection is via quasielastic charged-current scattering (QECC), \ie, $\nu_l + n \rightarrow p + l^-$ and $\bar{\nu}_l + p \rightarrow n + l^+$ ($l = e, \mu, \tau$), and via charged-current deep inelastic scattering (DIS), \ie, $\nu_l + N \rightarrow l^- + X$ and $\bar{\nu}_l + N \rightarrow l^+ + X$ ($l = e, \mu$), where $X$ represents final-state hadrons ($\nu_\tau$ DIS is suppressed due to the large tauon mass).  Electrons emit gamma rays by bremsstrahlung and $e^+e^-$ annihilation, which register as a fuzzy ring on the PMTs.  Muons emit Cherenkov light, which registers as a sharply defined ring.

Like its predecessor, T2K (Tokai-to-Kamioka)~\cite{T2K:2023smv}, T2HK will use the 2.5$^{\circ}$-off-axis JPARC neutrino beam~\cite{McDonald:2001mc}.  To produce it, JPARC fires a 1.3-MW beam of protons of 30~GeV onto a graphite target.  The resulting neutrino flux is narrow-band, ranges from a few~MeV to a few~GeV, and is expected to peak at 600~MeV, with most neutrinos in the 100--3000~MeV range.  As in DUNE, by changing the polarity of the focusing horns, T2HK can run in neutrino or antineutrino mode~\cite{T2K:2012bge}.  Following \Refe~\cite{Hyper-Kamiokande:2016srs}, we adopt a run time of 2.5 years in neutrino mode and 7.5 years in antineutrino mode, in accordance with the default 1:3 ratio planned for them.  This amounts to $2.7 \times 10^{22}$ protons-on-target per year and a net exposure of 2431 kton~MW~year.  To produce our results, we match the binned event spectra that we generate under standard oscillations with those of \Refe~\cite{Hyper-Kamiokande:2016srs}.
\subsection{Oscillation probabilities}
\label{sec:experiments_probabilities}
%
Appendix~\ref{app:evol_mix_param}, especially \figu{evol_mix_param} therein, shows in detail the effects of long-range interactions on the modified mixing parameters $\theta_{12}^m$, $\theta_{23}^m$, and $\theta_{13}^m$; here, we summarize them.  Differences in their behavior under the different symmetries stem from differences in the flavor structure of the new matter potential, $\mathbf{V}_{\alpha\beta}$ in \equ{hamiltonian_tot}.

The solar angle in matter rapidly approximates its maximum value of $\theta_{12}^m = 90^\circ$ already at a few GeV, for all symmetries. (We use this later, in Section~\ref{sec:results_stat_methods}, to justify why we neglect the effect on our forecasts of the uncertainty in its value in vacuum, $\theta_{12}$.)  For DUNE and T2HK, the mixing angles that drive the probabilities are the atmospheric angle, $\theta_{23}^m$, and the reactor angle, $\theta_{13}^m$.  Assuming $\theta_{12}^m = 90^\circ$, \Refe~\cite{Khatun:2018lzs} showed that the transition probabilities for $\nu_\mu \to \nu_e$ and $\bar{\nu}_\mu \to \bar{\nu}_e$ are $\propto \sin^2 \theta_{23}^m \sin^2 \theta_{13}^m$ and the survival probabilities for $\nu_\mu \to \nu_\mu$ and $\bar{\nu}_\mu \to 
\bar{\nu}_\mu$ are $\propto \sin^2 2\theta_{23}^m$, with a more nuanced dependence on $\theta_{13}^m$.  The deviation of $\theta_{23}^m$ relative to $\theta_{23}$ grows with energy, though there are differences depending on which symmetry is active: $\theta_{23}^m$ grows under $L_e-L_\tau$ and $L_\mu-L_\tau$, and shrinks under $L_e-L_\mu$.  The reactor angle in matter, $\theta_{13}^m$, grows appreciably with energy under $L_e-L_\mu$ and $L_e-L_\tau$, and falls to about $0^\circ$ under $L_\mu-L_\tau$.

Figure~\ref{fig:probabilities} shows the oscillation probabilities, \equ{osc_prob}, computed under the three symmetries in each of the four detection channels listed in Section~\ref{sec:experiments_overview}, for DUNE and T2HK.  To illustrate the effects of long-range interactions, we pick a relatively high value of the potential, $V_{\alpha\beta} = 1.3 \times 10^{-13}$~eV; later, when producing our results, we vary this value.  Via \equ{pot_total}, multiple combinations of $m_{\alpha\beta}^\prime$ and $G_{\alpha\beta}^\prime$ can yield this value of the potential, or any other.  Because the baseline for DUNE is longer than for T2HK, the effects of long-range interactions with underground matter on the probabilities in the former are more prominent than in the latter~\cite{Agarwalla:2021zfr}.  The effects are more clearly visible in the transition probabilities: the oscillation maxima shift to lower energies, due to a change in the effective mass-splitting $\Delta m_{31,m}^2$, in agreement with \Refe~\cite{Chatterjee:2015gta}, and the oscillation amplitudes grow, especially after the first maximum.  The effects are more prominent under $L_e-L_\tau$ because $\theta_{23}^m$ and $\theta_{13}^m$ are enhanced, whereas under $L_e-L_\mu$ and $L_\mu-L_\tau$ only one of them is; see Appendix~\ref{app:evol_mix_param} for details.  Naturally, for weaker potentials, the above effects are lessened. 
%
\subsection{Event rates}
\label{sec:experiments_event_rates}
\begin{table}[t!]
 \centering
 \begin{tabular}{ | c  *{6}{>{\centering\arraybackslash}p{2.2cm} |}}
 \hline
 \multicolumn{2}{|c|}{\multirow{3}{*}{Detector}}
 & \multicolumn{4}{c|}{Mean number of events (standard oscillations, NMO)} \\
          & & \multicolumn{2}{c|}{Appearance} 
          & \multicolumn{2}{c|}{Disappearance} \\
          & & $\nu$ mode & $\bar{\nu}$ mode & $\nu$ mode & $\bar{\nu}$ mode \\
 \hline
 \multirow{2}{*}{DUNE} & Signal &    1390 & 387 & 15574 & 8975 \\
                       & Bkg.   & 690  & 457  & 347   & 210   \\
 \hline
 \multirow{2}{*}{T2HK} & Signal & 1374 & 1166 & 10083 & 13905 \\
                       & Bkg.   & 802  & 991  & 1686  & 1769  \\
 \hline
 \end{tabular}
 \caption{\textbf{\textit{Mean number of signal and background events, summed over all background channels, expected in DUNE and T2HK after their full run times.}}  For this table, whose aim is illustrative only, we assume standard oscillations and normal mass ordering (NMO).  DUNE runs for 5 years in $\nu$ mode and 5 years in $\bar{\nu}$ mode.  T2HK runs for 2.5 years in $\nu$ mode and 7.5 years in $\bar{\nu}$ mode.  To compute the rates in this table, we fix the values of the standard mixing parameters to their benchmark values from \Refe~\cite{Capozzi:2021fjo}; see Table~\ref{tab:mix_param_benchmark}.  In the main text, to produce results, we also compute event rates in the presence of the new long-range neutrino interactions (not shown in this table).  In those cases, the relative sizes of the event rates in the different detection channels are roughly as in this table.  See Section~\ref{sec:experiments_event_rates} for details.
 }
 \label{tab:event}
\end{table}
%
\begin{figure}[t!]
 \centering
 \includegraphics[width=\linewidth]{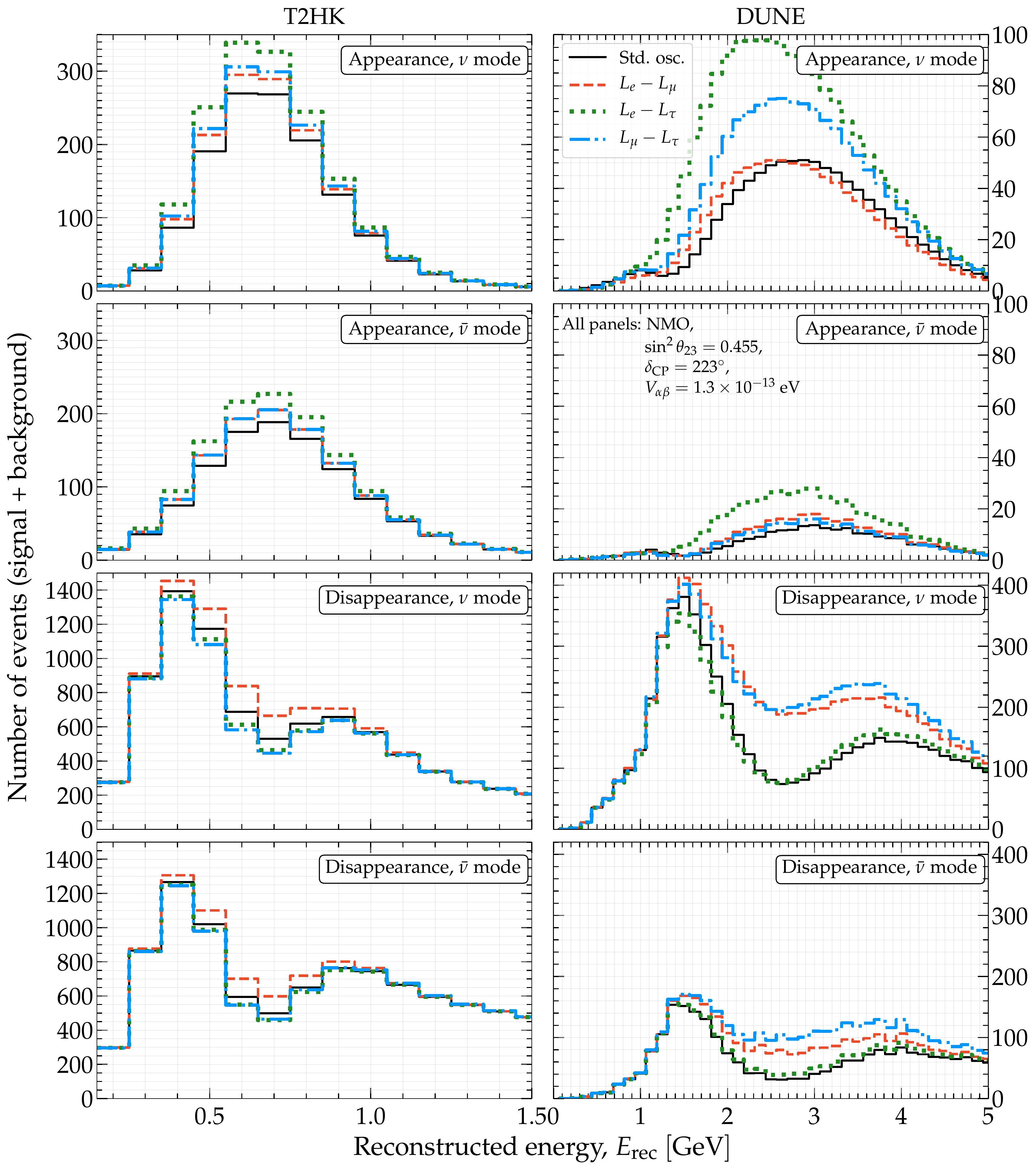}
 \caption{\textbf{\textit{Expected mean number of detected events in T2HK (left column) and DUNE (right column) under flavor-dependent long-range neutrino interactions.}}  The interactions are induced by the symmetry $L_e-L_\mu$, $L_e-L_\tau$, or $L_\mu-L_\tau$.  For T2HK, we use 2.5 years in $\nu$ mode and 7.5 years in $\bar{\nu}$ mode.  For DUNE, we use 5 years in $\nu$ mode and 5 years in $\bar{\nu}$ mode.  For this figure, we fix the long-range potential to $V_{\alpha\beta} = 1.3 \times 10^{-13}$~eV as illustration, and the standard mixing parameters to their benchmark values from Table~\ref{tab:mix_param_benchmark}.  See Section~\ref{sec:experiments_event_rates} for details.}
 \label{fig:event_rates}
\end{figure}
%
We compute event rates in DUNE and T2HK using {\sc GLoBES}~\cite{Huber:2004ka, Huber:2007ji},  extended with the {\sc snu} matrix-diagonalization library~\cite{Kopp:2006wp, Kopp:2007ne},  by modeling their technical design specifications~\cite{Hyper-Kamiokande:2018ofw, DUNE:2021cuw} of efficiency, operation times, and backgrounds.  Because we are interested in assessing the mean sensitivity of the experiments (Section~\ref{sec:results_stat_methods}), we compute only mean event rates and do not generate event spectra that include fluctuations from the mean rates.  We bin event rates in reconstructed energy, $E_{\rm rec}$, built from the detected secondaries born in neutrino interactions.  In both experiments, because the far detectors cannot distinguish between neutrinos and antineutrinos, there is irreducible contamination from ``wrong-sign'' events; we add it to the signal.

\medskip
\textbf{\textit{DUNE.--- }}We consider events with $E_{\rm rec}$ in the range 0--110~GeV, with 64 bins within 0--8~GeV, each 0.125~GeV wide, and 16 bins within 8--110~GeV, of varying widths. 
In the appearance channel, the signal is due to the charged-current (CC) interactions of $\nu_e$, in neutrino mode, and of $\bar{\nu}_e$, in antineutrino mode.  The background  consists of 
(i) the CC interactions of ``intrinsic'' $\nu_e$ and $\bar{\nu}_e$, \ie, those created as such that survive the flight to the detector (from $\nu_e \to \nu_e$ and $\bar{\nu}_e \to \bar{\nu}_e$); 
(ii) the CC interactions of $\nu_\mu$ and $\bar{\nu}_\mu$ whose final-state muons are misidentified as electrons (from $\nu_\mu \to \nu_\mu$ and $\bar{\nu}_\mu \to \bar{\nu}_\mu$); 
(iii) the CC interactions of $\nu_\tau$ and $\bar{\nu}_\tau$ (from $\nu_\mu \to \nu_\tau$ and $\bar{\nu}_\mu \to \bar{\nu}_\tau$); and 
(iv) the neutral-current (NC) interactions of neutrinos of all flavors.  
In the disappearance channel, the signal is due to the CC interactions of $\nu_\mu$, in neutrino mode, and of $\bar{\nu}_\mu$, in antineutrino mode.  The background consists of
(i) the CC interactions of $\nu_\tau$ and $\bar{\nu}_\tau$ (from $\nu_\mu \to \nu_\tau$ and $\bar{\nu}_\mu \to \bar{\nu}_\tau$); and 
(ii) the NC interactions of neutrinos of all flavors.   

\medskip

\textbf{\textit{T2HK.--- }}We consider events with $E_{\rm rec}$ in the range 0.1--3~GeV, with 29 bins, each 0.1~GeV wide.  In the appearance channel, the signal is due to the CC interactions of $\nu_e$, in neutrino mode, and of $\bar{\nu}_e$, in antineutrino mode.  The background consists of
(i) the CC interactions of intrinsic $\nu_e$ and $\bar{\nu}_e$  (from $\nu_e \to \nu_e$ and $\bar{\nu}_e \to \bar{\nu}_e$); 
(ii) the CC interactions of $\nu_\mu$ and $\bar{\nu}_\mu$ whose final-state muons produce fuzzy Cherenkov rings; and
(iii) the NC interactions of neutrinos of all flavors.
In the disappearance channel, the signal is due to the CC interactions of $\nu_\mu$, in neutrino mode, and of $\bar{\nu}_\mu$, in antineutrino mode.  The background consists of 
(i) the CC interactions of intrinsic $\nu_e$ and $\bar{\nu}_e$ (from $\nu_e \to \nu_e$ and $\bar{\nu}_e \to \bar{\nu}_e$); and 
(ii) the NC interactions of neutrinos of all flavors.

\medskip

Figure~\ref{fig:event_rates} shows the mean event-rate spectra under long-range interactions for each detection channel in DUNE and T2HK, including all the above backgrounds, and computed using the same illustrative value of the long-range potential $V_{\alpha\beta}$ as in \figu{probabilities}.  The event rates in T2HK are higher than in DUNE due to its larger size.  The shapes of the event spectra in \figu{event_rates} reflect those of the oscillation probabilities in \figu{probabilities}.  Long-range interactions affect each detection channel differently, but there are common features among them.  Broadly stated, in the appearance channels, they enhance the event rates relative to the standard-oscillations rates (with the exception of $L_e-L_\mu$ in neutrino mode for DUNE).  For DUNE, additionally, they slightly shift the event rates to lower energies, reflecting the shift in the oscillation maxima.  In the disappearance channels, the effect of long-range interactions is more nuanced; the event rate is enhanced or reduced depending on the symmetry and the energy.  The above features in the event spectra hold for other values of the potential, though, naturally, their prominence varies depending on the value.

Table~\ref{tab:event} shows the mean expected number of signal and background events for each detection channel, assuming standard oscillations.  In all channels, the signal is dominant.  In T2HK, unlike DUNE, neutrino and antineutrino event rates are comparable, due to the 1:3 ratio between run times in neutrino and antineutrino modes that compensates for the smaller antineutrino cross sections.  In DUNE, neutrino event rates are higher than T2HK due to its longer run time in neutrino mode.  These general features of the event rates hold also in the presence of long-range interactions.  

Below, we show how the above features grant DUNE and T2HK the capability to probe long-range interactions, and how they organically complement each other.
%
\section{Projected constraints and discovery potential}
\label{sec:results}
\subsection{Statistical methods}
\label{sec:results_stat_methods}
We forecast the capability of DUNE and T2HK to probe long-range interactions that stems from the modification of the oscillation probabilities (Section~\ref{sec:experiments_probabilities}), based on the detailed computation of event rates outlined above (Section~\ref{sec:experiments_event_rates}).  Our forecasts are two-fold: we forecast {\it constraints} on long-range interactions --- on the long-range matter potential and ultimately on the mediator mass and coupling --- assuming that no evidence for them is found, and we forecast prospects of {\it discovering} them and measuring their parameter values.

We study each symmetry, $L_e-L_\mu$, $L_e-L_\tau$, and $L_\mu-L_\tau$, separately.  For a given symmetry, we generate two sets of event spectra, including signal plus backgrounds, for each of the four detection channels of T2HK and DUNE (Section~\ref{sec:experiments_event_rates}): a ``true'' spectrum, which we take to be the observed spectrum, and a set of ``test'' spectra, generated for test values of the parameters, that we compare against it.  When forecasting constraints, in Section~\ref{sec:results_constraints}, we compute the true spectrum fixing the true value of the potential to be $V_{\alpha\beta}^{\rm true} = 0$, which corresponds to standard oscillations.  When forecasting discovery prospects, in Section~\ref{sec:results_discovery}, we compute the true spectrum fixing $V_{\alpha\beta}^{\rm true}$ to a specific nonzero choice.  We expand on this below. 

To compare true and test event spectra, we follow \Refes~\cite{Baker:1983tu, Cowan:2010js, Blennow:2013oma} and adopt a Poissonian $\chi^2$ function. For each experiment $e = \{{\rm T2HK},~{\rm DUNE}\}$, and for each detection channel $c = \{{\rm app}~\nu,~{\rm app}~\bar{\nu},~{\rm disapp}~\nu,~{\rm disapp}~\bar{\nu}\}$, this is
\begin{eqnarray}
 \chi_{e,c}^{2}
 (V_{\alpha\beta}, \boldsymbol{\theta}, o)
 =
 &&
 \underset{\left\{\xi_{s}, \{\xi_{b, c, k}\}\right\}}{\mathrm{min}} 
 \left\{
 2\sum^{N_e}_{i=1}
 \left[
 N_{e, c, i}^{{\rm test}}
 (V_{\alpha\beta}, \boldsymbol{\theta}, o, \xi_s, \{\xi_{b,c,k}\})
 \right. \right.
 \nonumber \\
 &&
 \left. \left.
 -
 N_{e, c, i}^{{\rm true}}
 \left( 
 1
 +
 \ln
 \frac{N_{e, c, i}^{{\rm test}}
 (V_{\alpha\beta}, \boldsymbol{\theta}, o, \xi_s, 
 \{\xi_{b, c, k}\})}
 {N_{e, c, i}^{{\rm true}}}
 \right)
 \right]
 +
 \xi^{2}_{s}
 +
 \sum_k \xi^{2}_{b, c, k} 
 \right\} \;,
\end{eqnarray}
where  $N_{e, c, i}^{{\rm true}}$ and $N_{e, c, i}^{{\rm test}}$ are the true and test event rates in the $i$-th bin of $E_{\rm rec}$, $N_e$ is the number of bins of $E_{\rm rec}$ (Section~\ref{sec:experiments_event_rates}), $\boldsymbol{\theta} \equiv \{ \sin^2\theta_{23}, \delta_{\rm CP}, \vert \Delta m_{31}^2\vert \}$ are the test values of the most relevant mixing parameters (more on this later), $o = \{ \text{NMO, IMO} \}$ is the test mass ordering, and $\xi_{s}$ and $\xi_{b, c, k}$ are, respectively, pull terms for the systematic uncertainties on the signal and the $k$-th background contribution to detection channel $c$, from the list of contributions in Section~\ref{sec:experiments_event_rates}.  The pull terms have the same values in neutrino and antineutrino mode, and are uncorrelated with one another.  The true number of events is
\begin{equation}
 N_{e, c, i}^{{\rm true}}
 = 
 N_{e, c, i}^{s, {\rm true}}
 + 
 N_{e, c, i}^{b, {\rm true}} \;,
\end{equation}
where $N_{e, c, i}^{s, {\rm true}}$ and $N_{e, c, i}^{b, {\rm true}}$ are, respectively, the number of signal ($s$) and background ($b$) events, summed over all channels, computed using the true values of the mixing parameters, mass ordering, and potential.
The test number of events is
\begin{equation}
 \label{equ:num_test}
 N_{e, c, i}^{{\rm test}}
 (V_{\alpha\beta}, \boldsymbol{\theta}, o, \xi_s, \left\{\xi_{b,c,k}\right\})
 =
 N^s_{e,c,i}(V_{\alpha\beta}, \boldsymbol{\theta}, o)
 (1+\pi_{e,c}^s\xi_s)
 +
 \sum_k
 N^{b}_{e,c,k,i}(\boldsymbol{\theta}, o)
 \left(
 1+\pi_{e,c,k}^b\xi_{b,c,k}
 \right) \;,
\end{equation}
where $\pi_{e,c}^s$ and $\pi_{e,c,k}^b$ are normalization errors on the signal and background rates, which lie between 2\% and 10\%; see Table~\ref{tab:normalization_err} in Appendix~\ref{Appendix:4} for their  values, taken from \Refes~\cite{Hyper-Kamiokande:2018ofw, DUNE:2021cuw}.  The background rates do not vary significantly upon changing the mass ordering.

For T2HK or DUNE, separately or together, we compute the total $\chi^2$ by adding the contributions of all the detection channels, \ie,
\begin{eqnarray}
 \label{equ:chi2_dune}
 \chi_{\rm DUNE}^{2}
 (V_{\alpha\beta}, \boldsymbol{\theta}, o)
 &=&
 \sum_c \chi_{{\rm DUNE}, c}^{2}
 (V_{\alpha\beta}, \boldsymbol{\theta}, o)
 \;, \\
 \label{equ:chi2_t2hk}
 \chi_{\rm T2HK}^{2}
 (V_{\alpha\beta}, \boldsymbol{\theta}, o)
 &=&
 \sum_c \chi_{{\rm T2HK}, c}^{2}
 (V_{\alpha\beta}, \boldsymbol{\theta}, o) 
 \;, \\
 \label{equ:chi2_dune_t2hk}
 \chi^2_{{\rm DUNE}+{\rm T2HK}}
 (V_{\alpha\beta}, \boldsymbol{\theta}, o)
 &=&
 \chi^2_{\rm DUNE}
 (V_{\alpha\beta}, \boldsymbol{\theta}, o)
 +
 \chi^2_{\rm T2HK}
 (V_{\alpha\beta}, \boldsymbol{\theta}, o)
 \;, 
\end{eqnarray}
and we take the contributions of different channels to be uncorrelated.  

We report sensitivity by comparing the minimum value of the $\chi^2$ function, $\chi_{e, \rm min}^2$, reached when it is evaluated at the true values of the parameters, $V_{\alpha\beta}^{\rm true}$, $\boldsymbol{\theta}^{\rm true}$, and $o^{\rm true}$, against the value of $\chi^2$ evaluated at test values of the parameters.  We treat $\boldsymbol{\theta}$ and $o$ as nuance parameters and profile over them (more on this below).  For instance, for DUNE,
\begin{equation}
 \label{equ:delta_chi2_dune}
 \Delta\chi^2_{\rm DUNE}(V_{\alpha\beta})
 =
 \underset
 {\{ \boldsymbol{\theta}, o\}}{\mathrm{min}} 
 \left[
 \chi_{\rm DUNE}^{2}
 (V_{\alpha\beta}, \boldsymbol{\theta}, o) 
 -
 \chi_{\rm DUNE, min}^{2}
 \right]
 \;,
\end{equation}
and similarly for T2HK and DUNE+T2HK.  In the main text, we fix $\boldsymbol{\theta}^{\rm true}$ to its present-day best-fit value under NMO, from \Refe~\cite{Capozzi:2021fjo} (Table~\ref{tab:mix_param_benchmark}) and $o^{\rm true}$ also to NMO; we fix them to inverted mass ordering in Appendix~\ref{app:imo}.  When reporting {\it constraints} on long-range interactions, we set $V_{\alpha\beta}^{\rm true} = 0$ and extract from $\Delta \chi_{\rm DUNE}^2$, $\Delta \chi_{\rm T2HK}^2$, and $\Delta \chi_{\rm DUNE+T2HK}^2$ the upper limits on the inferred value of $V_{\alpha\beta}$, for 1~degree of freedom (d.o.f).  When reporting {\it discovery potential}, we fix $V_{\alpha\beta}^{\rm true}$ to a nonzero illustrative value and report the inferred range of values of $V_{\alpha\beta}$, again for 1~d.o.f.  When reporting discovery, we also study the correlation between $V_{\alpha\beta}$ and $\dcp$ or $\sin^2\theta_{23}$.  In those cases, we use instead, respectively,
\begin{eqnarray}
 \label{equ:delta_chi2_2dof_dune_vs_dcp}
 \Delta\chi^2_{\rm DUNE}
 (V_{\alpha\beta}, \dcp)
 &=&
 \underset
 {\{ \sin^2 \theta_{23}, \left\vert \Delta m_{31}^2 \right\vert, o\}}{\mathrm{min}} 
 \left[
 \chi_{\rm DUNE}^{2}
 (V_{\alpha\beta}, \boldsymbol{\theta}, o) 
 -
 \chi_{\rm DUNE, min}^{2}
 \right]
 \;, \\
 \label{equ:delta_chi2_2dof_dune_vs_th23}
 \Delta\chi^2_{\rm DUNE}
 (V_{\alpha\beta}, \sin^2 \theta_{23})
 &=&
 \underset
 {\{ \dcp, \left\vert \Delta m_{31}^2 \right\vert, o\}}{\mathrm{min}} 
 \left[
 \chi_{\rm DUNE}^{2}
 (V_{\alpha\beta}, \boldsymbol{\theta}, o) 
 -
 \chi_{\rm DUNE, min}^{2}
 \right]
 \;,
\end{eqnarray}
and similarly for T2HK and DUNE + T2HK, and we show allowed regions for 2~d.o.f.  After placing bounds on $V_{\alpha\beta}$, we translate them into bounds on $G_{\alpha\beta}^\prime$ for varying $m_{\alpha\beta}^\prime$ by means of the definition of the long-range matter potential, \equ{pot_total}. 

When profiling, we minimize the $\Delta\chi^2$ functions above with respect to $\sin^{2}\theta_{23}$, $\delta_{\mathrm{CP}}$, and $\vert \Delta m^{2}_{31} \vert$ by varying them within their $3\sigma$ allowed ranges from \Refe~\cite{Capozzi:2021fjo}; see Table~\ref{tab:mix_param_benchmark}.  We do not include correlations between them, since these are expected to weaken in coming years (see, \eg, \Refe~\cite{Song:2020nfh}), nor do we include pull terms on the mixing parameters in the test-statistic, in order to be conservative.  We keep $\theta_{13}$ and $\theta_{12}$ fixed at their present-day best-fit values~\cite{Capozzi:2021fjo}.  For $\theta_{13}$, the precision that Daya Bay has achieved, of 2.8\%~\cite{DayaBay:2022orm}, is not expected to be improved upon by upcoming experiments.  For $\theta_{12}$, whose present-day uncertainty is of 4.5\%, we expect only weak sensitivity in the oscillation probabilities (see Section~\ref{sec:experiments_probabilities}), so fixing its value is a justified approximation.  For the neutrino mass ordering, we adopt a simplified approach where switching from NMO to IMO amounts only to flipping the sign of $\Delta m_{31}^2$ to make it negative.  This approach is motivated by the fact that the present-day 3$\sigma$ allowed ranges of the mixing parameters are similar in the NMO and IMO~\cite{Capozzi:2021fjo, Esteban:2020cvm, NuFIT}, except for $\dcp$ and $\Delta m_{31}^2$.  (Further, in the next decade, DUNE is expected to determine the mass ordering, though, admittedly, non-standard oscillations like those induced by long-range interactions may confound this~\cite{Chatterjee:2015gta}.) 

Below, we compute the test-statistics by varying the long-range matter potential in the range $10^{-15} \leq V_{\alpha\beta}/{\rm eV} \leq 3 \times 10^{-13}$, where its effects are potentially visible in DUNE and T2HK.  This range is wide enough to comfortably place constraints or make measurements with significant statistical confidence.
\subsection{Projected constraints on long-range interactions}
\label{sec:results_constraints}
%
\begin{figure}[t!]
 \centering
 \includegraphics[width=\linewidth]{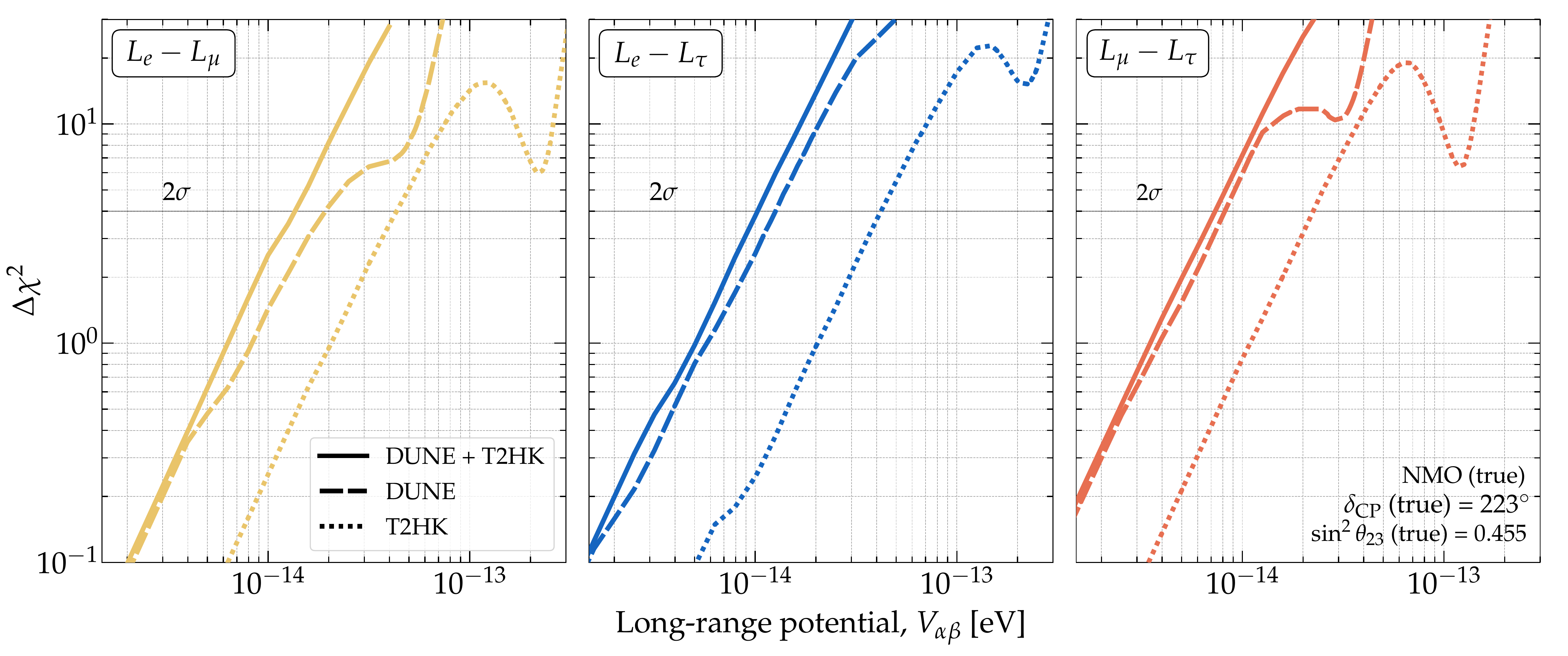}
 \caption{\textbf{\textit{Projected test-statistic used to constrain the long-range matter potentials $V_{e\mu}$, $V_{e\tau}$, and $V_{\mu\tau}$, using DUNE, T2HK, and their combination.}}  The $\Delta \chi^2$ function is \equ{delta_chi2_dune} and similar ones, assuming $V_{\alpha\beta}^{\rm true} = 0$ for the true value of the potentials.  We profile over the values of the most relevant standard mixing parameters and over the neutrino mass ordering; see Table~\ref{tab:mix_param_benchmark}.  See Table~\ref{tab:upper_limits_potential} for the resulting upper limits on the potentials and \figu{constraints_g_vs_m} for the corresponding constraints on the mass and coupling of the new mediator.  \textit{Combining DUNE and T2HK not only provides sensitivity to lower values of the potential, but also removes degeneracies in the test-statistics that would otherwise weaken the sensitivity.} See Sections~\ref{sec:results_stat_methods} and \ref{sec:results_constraints} for details.}
 \label{fig:test_statistics}
\end{figure}
%
Figure~\ref{fig:test_statistics} shows how the test-statistics for constraints, \eg, \equ{delta_chi2_dune}, vary with the potential, for the three symmetries and for DUNE, T2HK, and their combination.  As expected, they are smallest closer to the true value of the potential, $V_{\alpha\beta}^{\rm true} = 0$, and grow as they move away from it.  At high values of $V_{\alpha\beta}$, the test-statistics for DUNE and T2HK dip, reflecting a loss of sensitivity due to $V_{\alpha\beta}$ being degenerate with $\theta_{23}$ and $\dcp$. Appendix~\ref{Appendix:2} expands on this.  Combining DUNE and T2HK removes the dips: T2HK lifts the degeneracies due to $\theta_{23}$ and $\dcp$, while DUNE fixes the mass ordering, \ie, the sign of $\Delta m_{31}^2$.  Thus, our results reveal novel insight: \textbf{\textit{the interplay of DUNE and T2HK facilitates degeneracy-free constraints on flavor-dependent long-range neutrino interactions.}}

\begin{figure}[t!]
 \centering
 \includegraphics[width=\linewidth]{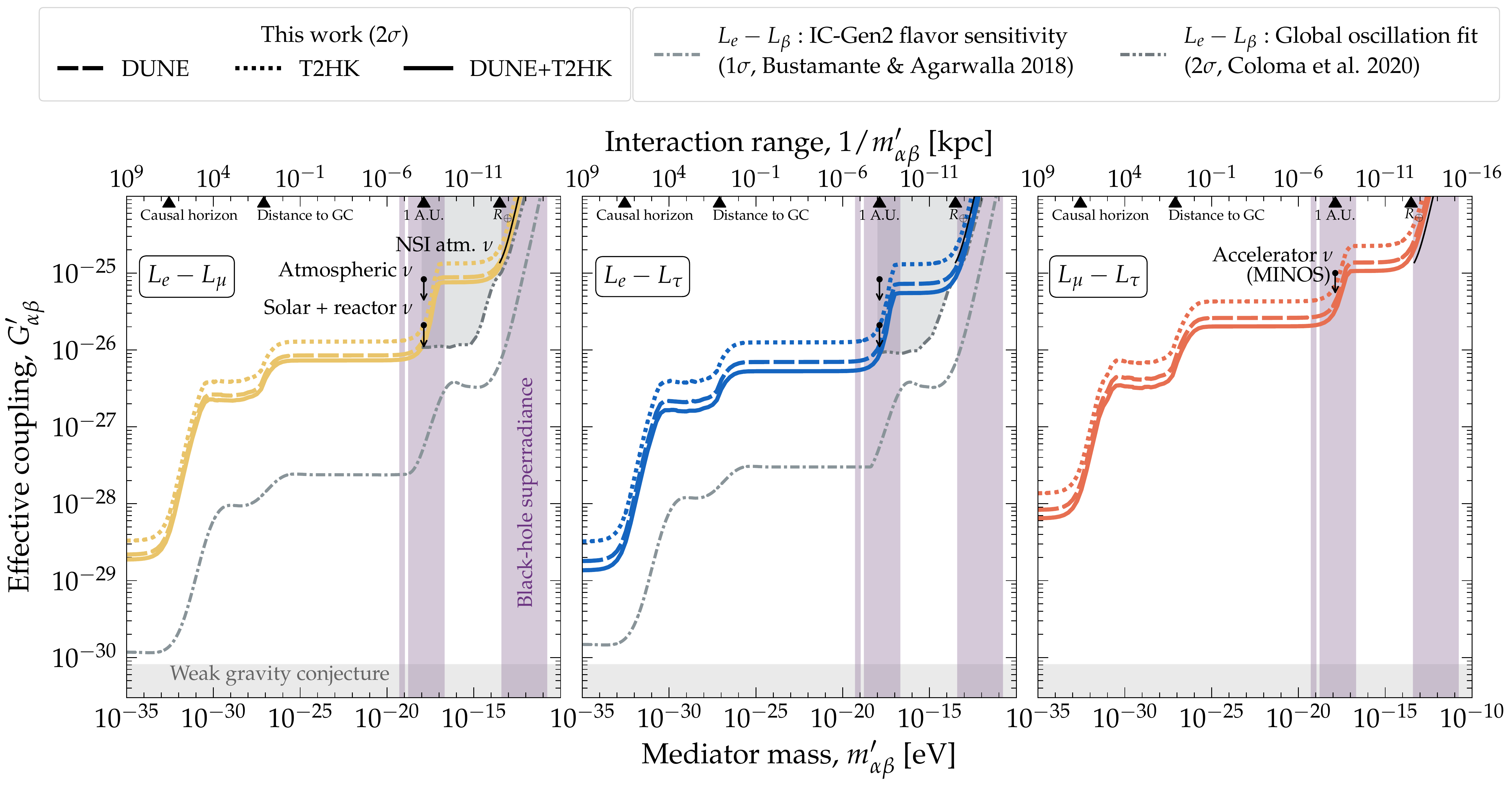}
 \caption{\textbf{\textit{Projected upper limits on the effective coupling, $G_{\alpha\beta}^\prime$ (Eq.~(\ref{equ:Gab})), of the new boson, $Z_{\alpha\beta}^\prime$, with mass $m_{\alpha\beta}^\prime$, that mediates flavor-dependent long-range neutrino interactions, using DUNE, T2HK, and their combination.}}  Same as \figu{moneyplot}, but now showing also limits using DUNE and T2HK separately.  {\it Left:} For neutrino-electron interactions under the $L_e-L_\mu$ symmetry.  {\it Center:} For neutrino-electron interactions under $L_e-L_\tau$.  {\it Right:} For neutrino-neutron interactions under $L_\mu-L_\tau$; existing limits from accelerator neutrinos in MINOS (95\% C.L.) are from \Refe~\cite{Heeck:2010pg}, and for non-standard interactions (NSI) we follow Appendix C in Ref.~\cite{Agarwalla:2023sng}. See Section~\ref{sec:results_constraints} for details.}
 \label{fig:constraints_g_vs_m}
\end{figure}

Table~\ref{tab:upper_limits_potential} shows the resulting upper limits on the potential.  They are strongest for $V_{\mu\tau}$, followed by $V_{e\tau}$ and then $V_{e\mu}$.  For DUNE, the limits are driven predominantly by the runs in neutrino mode, which contribute most of the total event rate; see Table~\ref{tab:event} and \figu{event_rates}.  For T2HK, the runs in neutrino and antineutrino modes contribute comparably.  For $L_\mu-L_\tau$, the limits on $V_{\mu\tau}$ are strongest because long-range interactions affect mainly the disappearance probabilities, $\nu_\mu \to \nu_\mu$ and $\bar{\nu}_\mu \to \bar{\nu}_\mu$ (see \figu{probabilities}), whose associated  disappearance detection channels have high event rates (see \figu{event_rates}), making deviations from standard oscillations easier to spot.  For $L_e-L_\tau$, long-range interactions enhance instead the appearance probabilities, $\nu_\mu \to \nu_e$ and $\bar{\nu}_\mu \to \bar{\nu}_e$, but the appearance detection channels have lower rates, so the limits on $V_{e\tau}$ are weaker.  For $L_e-L_\mu$, long-range interactions affect both the appearance probabilities --- though less so than under the other two symmetries --- and the disappearance probabilities --- though less so than under the $L_\mu-L_\tau$ symmetry; as a result, the limits on $V_{e\mu}$ are the weakest.  
%
\begin{table}[b!]
\centering
 \begin{tabular}{ | c | *{4}{>{\centering\arraybackslash}p{2.25cm} |}}
 \hline
 \multirow{2}{*}{Detector} &
 \multicolumn{3}{c|}{Upper limits ($2\sigma$) on  potential [$10^{-14}$~eV]} \\
  & $V_{e\mu}$ & $V_{e\tau}$ & $V_{\mu\tau}$ \\
 \hline
 DUNE        & $1.9$ & $1.3$ & $0.82$ \\
 T2HK        & $4.4$ & $4.2$ & $2.2$  \\
 DUNE + T2HK & $1.4$ & $1.0$ & $0.73$ \\
 \hline
 \end{tabular}
 \caption{\textbf{\textit{Projected upper limits (2$\sigma$) on the long-range matter potentials $V_{e\mu}$, $V_{e\tau}$, and $V_{\mu\tau}$, using DUNE, T2HK, and their combination.}}  See \figu{test_statistics} for the test-statistics from whence they originate and \figu{constraints_g_vs_m} for constraints on the mass and coupling of the associated mediator. See Sections~\ref{sec:results_stat_methods} and \ref{sec:results_constraints} for details.}
 \label{tab:upper_limits_potential}
\end{table}
%
Figure~\ref{fig:constraints_g_vs_m} (also \figu{moneyplot}) shows the corresponding upper limits on $G_{\alpha\beta}^\prime$ (refer to Eq.~(\ref{equ:Gab})) for varying $m_{\alpha\beta}^\prime$, translated from the upper limits on $V_{\alpha\beta}$ in Table~\ref{tab:upper_limits_potential} via the definition of the potential, \equ{pot_total}.  Each curve in \figu{constraints_g_vs_m} is an isocontour of potential that saturates each of the upper limits in Table~\ref{tab:upper_limits_potential}.  The curves show step-like transitions at various values of $m_{\alpha\beta}^\prime$: as explained in \Refe~\cite{Bustamante:2018mzu} (see also Section~\ref{sec:lri_potential}), each transition reflects the interaction range becoming long enough for a new source of electrons or neutrons to contribute to the total potential, \equ{pot_total}.  For $m_{\alpha\beta}^\prime \sim 10^{-18}$--$10^{-10}$~eV, the Earth and the Moon dominate the upper limits; for $m_{\alpha\beta}^\prime \lesssim 10^{-18}$~eV, the Sun dominates; for $m_{\alpha\beta}^\prime \lesssim 10^{-27}$~eV, the Milky Way dominates; and, for $m_{\alpha\beta}^\prime \lesssim 10^{-33}$~eV, cosmological electrons and neutrinos dominate.  

Down to $m_{\alpha\beta}^\prime \sim 10^{-18}$~eV, where direct limits on flavor-dependent long-range neutrino interactions exist,  our projected limits improve on existing ones that use atmospheric neutrinos~\cite{Joshipura:2003jh}, are comparable to limits that use solar, reactor~\cite{Bandyopadhyay:2006uh}, and accelerator neutrinos~\cite{Heeck:2010pg}, and to limits culled from non-standard interactions~\cite{Super-Kamiokande:2011dam, Ohlsson:2012kf, Gonzalez-Garcia:2013usa}, but are weaker than limits from a global fit to oscillation data~\cite{Coloma:2020gfv}.  
%

Below $m_{\alpha\beta}^\prime \sim 10^{-18}$~eV, our projected limits tread into a largely unexplored range.  To our knowledge, the only constraints that exist there, other than the indirect, tentative one from the weak gravity conjecture~\cite{Arkani-Hamed:2006emk}, are from measurements of the flavor composition of high-energy astrophysical neutrinos in the IceCube neutrino telescope, from \Refe~\cite{Bustamante:2018mzu}, which, however, were only produced at the $1\sigma$ level as a proof of principle of the sensitivity.  A recent recalculation~\cite{Agarwalla:2023sng} found comparable results at higher statistical significance, using more solid statistical methods.  Ideally, the sensitivity that could be reaped from high-energy astrophysical neutrinos is unmatched due to them having energies in the TeV--PeV range --- which enhances the possible contribution of long-range interactions relative to standard oscillations --- and to the fact that neutrinos of all flavors are detected.  However, it is presently downplayed by large astrophysical uncertainties, limited event rates, and the difficulty in measuring the flavor composition in neutrino telescopes.  These issues will likely be surmounted in the future~\cite{Ackermann:2019cxh, Arguelles:2019rbn, Ackermann:2022rqc}.  For now, \figu{constraints_g_vs_m} shows the projected proof-of-principle sensitivity of the envisioned IceCube-Gen2 upgrade~\cite{IceCube-Gen2:2020qha}, from \Refe~\cite{Bustamante:2018mzu}.  Our limits from DUNE and T2HK improve on it significantly due to high event rates and well-characterized neutrino beams.
\subsection{Discovering subdominant long-range interactions}
\label{sec:results_discovery}
\begin{figure}
 \centering
 \includegraphics[width=\linewidth]{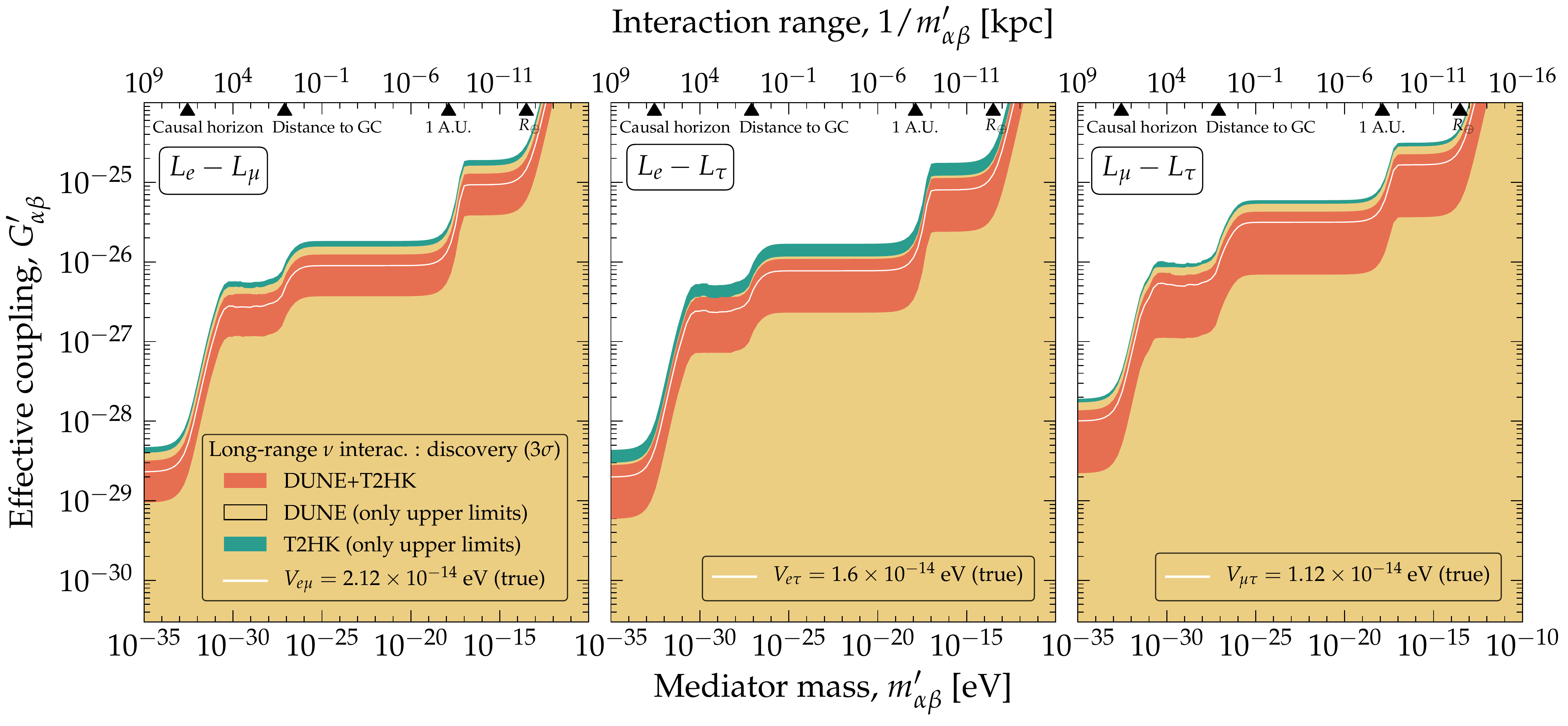}
 \caption{\textbf{\textit{Projected discovery potential of flavor-dependent long-range neutrino interactions.}}  We show allowed ranges ($3\sigma$) of the effective coupling, $G_{\alpha\beta}^\prime$ (Eq.~(\ref{equ:Gab})), of the new boson, $Z_{\alpha\beta}^\prime$, with mass $m_{\alpha\beta}^\prime$, that mediates the interactions, using DUNE, T2HK, and their combination.  The $\Delta \chi^2$ function is \equ{delta_chi2_dune} and similar ones, fixing the true value of the long-range potential, $V_{\alpha\beta}^{\rm true}$, at test values chosen to make the long-range interactions subdominant.  Like in \figu{constraints_g_vs_m}, we either fix or profile over the standard mixing parameters and the neutrino mass ordering; see Table~\ref{tab:mix_param_benchmark}.  See related Figs.~\ref{fig:ranges_dcp_V} and \ref{fig:ranges_th23_V}.  \textit{DUNE or T2HK may not be able to discover long-range interactions separately, but their combination may.} See Sections~\ref{sec:results_stat_methods} and \ref{sec:results_discovery} for details.}
 \label{fig:g_vs_m_discovery}
\end{figure}
%
\begin{figure}[t!]
 \includegraphics[width=\linewidth]{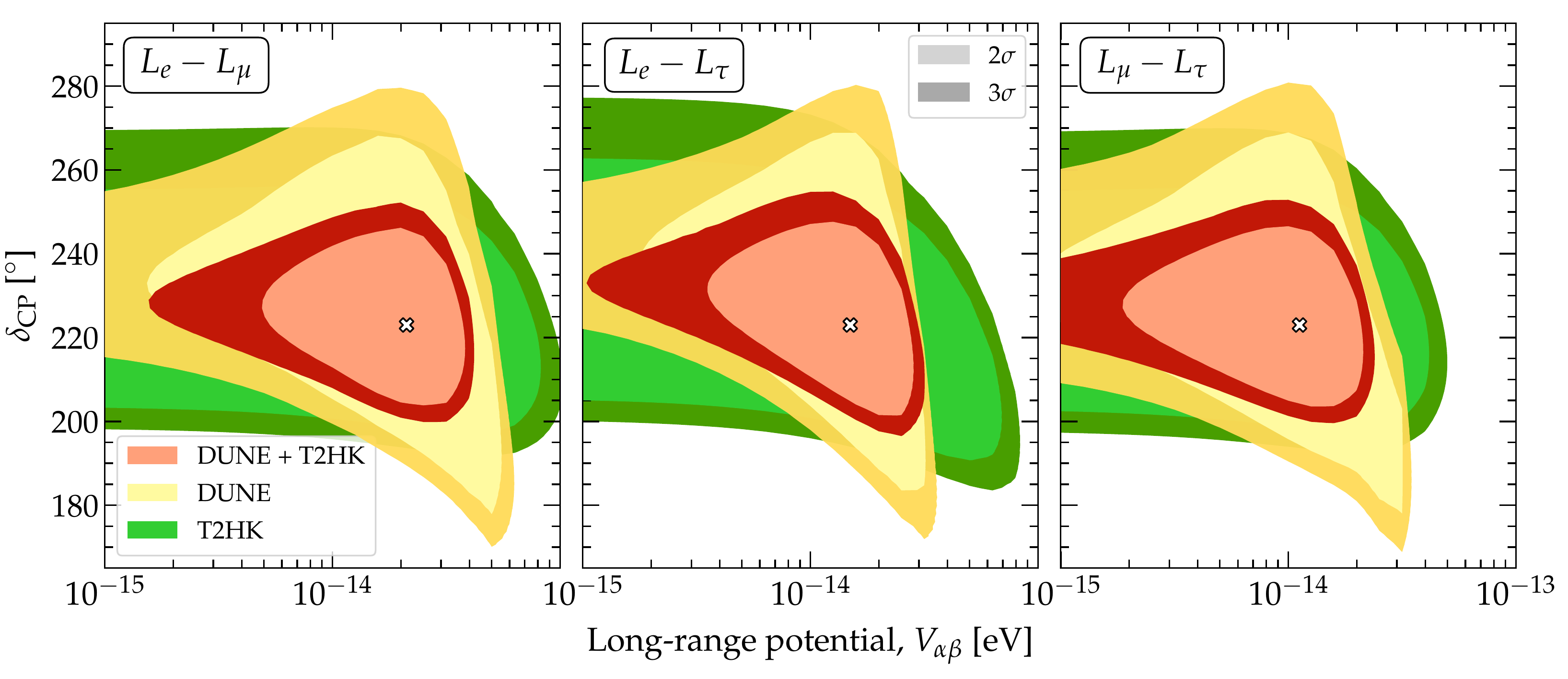}
 \caption{\textbf{\textit{Allowed regions of the long-range potential, $V_{\alpha\beta}$, and the CP-violating phase, $\dcp$.}}  The true values of the potentials are the same as in \figu{g_vs_m_discovery}.  The test-statistic is profiled over $\sin^2\theta_{23}$, $\lvert \Delta m^2_{31} \rvert$, and the mass ordering; see, \eg, \equ{delta_chi2_2dof_dune_vs_dcp}.  See Sections~\ref{sec:results_stat_methods} and \ref{sec:results_discovery} for details.}
 \label{fig:ranges_dcp_V}
\end{figure}
%
Figure~\ref{fig:g_vs_m_discovery} shows the inferred allowed ranges of the coupling, for varying mediator mass, for three illustrative choices of the true value of the long-range potential, one for each symmetry: $V_{e\mu}^{\rm true}$ = $2.12 \times 10^{-14}$~eV, $V_{e\tau}^{\rm true}$ = $1.6 \times 10^{-14}$~eV, and $V_{\mu\tau}^{\rm true}$ = $1.12 \times 10^{-14}$~eV.  They represent long-range interactions that are subdominant to standard oscillations; see Section~\ref{sec:lri_osc_prob}.   To generate \figu{g_vs_m_discovery}, first we compute the allowed ranges of $V_{\alpha\beta}$ using the discovery test-statistics, \eg, \equ{delta_chi2_dune} with $V_{\alpha\beta}^{\rm true}$ fixed to the above illustrative choices, and then we use \equ{pot_total} to translate those into allowed ranges of $G_{\alpha\beta}^\prime$ for different values of $m_{\alpha\beta}^\prime$.  We present results for a modest discovery significance of $3\sigma$.

Figure~\ref{fig:g_vs_m_discovery} shows that DUNE and T2HK, by themselves, can only place upper limits on $G_{\alpha\beta}^\prime$.  Thus, our discovery forecasts also reveal novel insight: \textbf{\textit{DUNE and T2HK, by themselves, may be unable to discover subdominant long-range interactions, but their combined action may.}}  For the illustrative choices of the potentials, the allowed $3\sigma$ ranges combining DUNE and T2HK are $V_{e\mu} \in [3.62 \times 10^{-15}, 4.04 \times 10^{-14}]$~eV, $V_{e\tau} \in [1.41 \times 10^{-15}, 3.13 \times 10^{-14}]$~eV, and $V_{\mu\tau} \in [5.87 \times 10^{-16}, 2.24 \times 10^{-14}]$~eV, implying a relative measurement uncertainty of 90\%--100\%.  For larger values of the true potential, discovery claims should be stronger and the uncertainty in its measurement should shrink.

\begin{figure}[t!]
 \includegraphics[width=\linewidth]{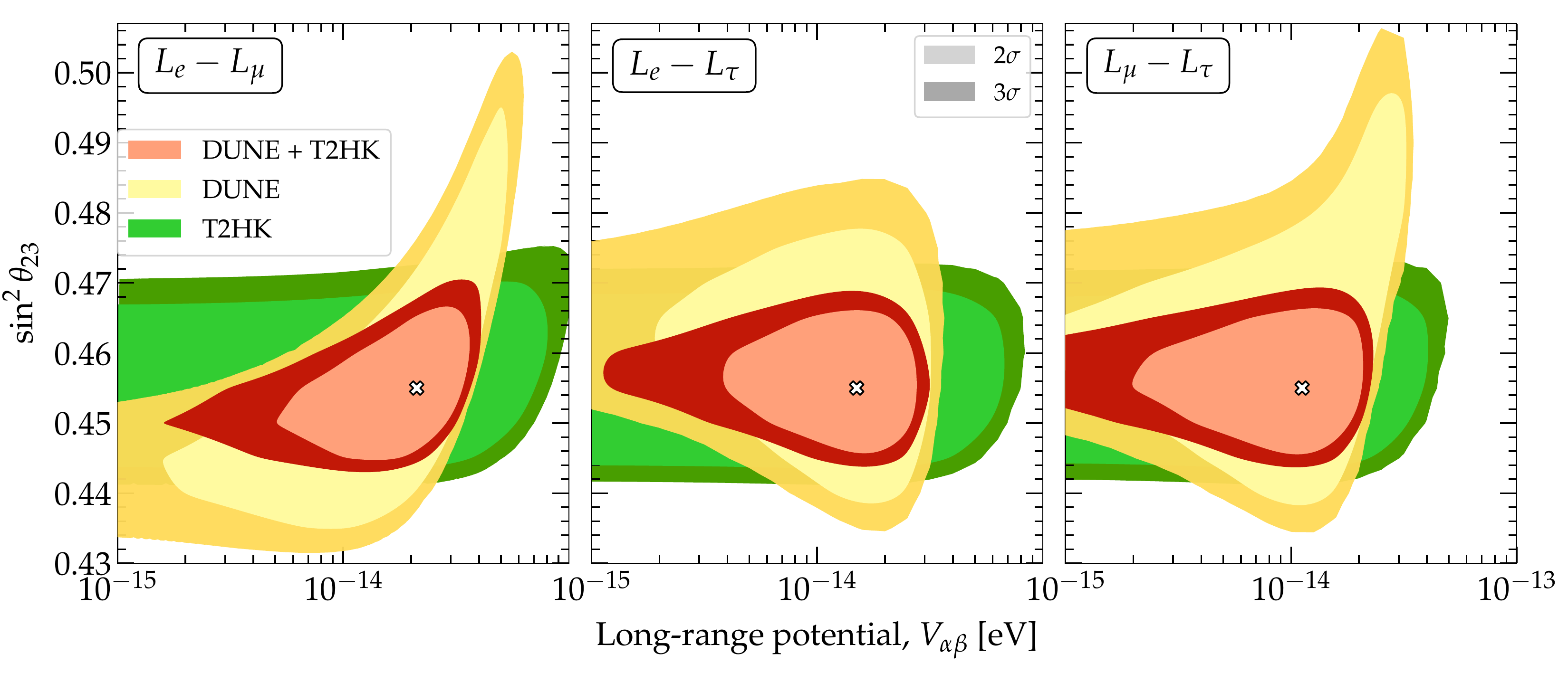}
 \caption{\textbf{\textit{Allowed regions of the long-range potential, $V_{\alpha\beta}$, and the atmospheric mixing angle, $\sin^2\theta_{23}$.}}  Same as \figu{ranges_dcp_V}, but profiling the test-statistic instead over $\dcp$, $\lvert\Delta m_{31}^2\rvert$, and the mass ordering; see, \eg, \equ{delta_chi2_2dof_dune_vs_th23}.  See Sections~\ref{sec:results_stat_methods} and \ref{sec:results_discovery} for details.}
 \label{fig:ranges_th23_V}
\end{figure}

Figures~\ref{fig:ranges_dcp_V} and \ref{fig:ranges_th23_V} reveal that the reason behind the difficulty of T2HK and DUNE to discover subdominant long-range interactions by themselves are the uncertainties in $\dcp$, $\theta_{23}$, and the neutrino mass ordering.  Appendix~\ref{Appendix:2} shows this in detail; here we summarize.  On the one hand, in T2HK, the shorter baseline provides less contamination from fake CP violation induced by SM matter effects and, therefore, higher precision in measuring $\dcp$~\cite{Ballett:2016daj, Bernabeu:2018twl, Bernabeu:2018use, King:2020ydu, Agarwalla:2022xdo}, while the high event rates in the disappearance channels provide high precision on $\theta_{23}$.  However, at the same time, the shorter baseline reduces the sensitivity to the mass ordering and provides a shorter neutrino travel time during which long-range interactions may act.  On the other hand, in DUNE, the longer baseline helps to pin down the mass ordering, but introduces more contamination from fake CP violation, which degrades the sensitivity to $\dcp$ compared to T2HK.  Also, in the presence of long-range interactions, DUNE can measure $\theta_{23}$ significantly less precisely than T2HK.  
 
Thus, combining DUNE and T2HK improves the sensitivity with which $\dcp$, $\theta_{23}$, and the mass ordering can be measured, weakens the degeneracies between them and the long-range potential, and allows for its measurement at a high statistical significance.
%
%
\section{Summary and outlook}
\label{sec:conclusions}

Neutrinos are powerful probes of new physics.  Extant uncertainties in their properties leave room for the conceivable possibility that they experience interactions with matter beyond weak ones.  Discovering them would not only further our view of neutrino physics, but also represent striking evidence of physics beyond the Standard Model.  In the 2030s, the next-generation long-baseline neutrino experiments DUNE and T2HK may provide us with an opportunity to look for new neutrino interactions more incisively than ever before, thanks to high event rates and well-characterized neutrino beams.  We have forecast their reach.  Because we use detailed simulations of the detectors, including efficiencies, run times, and backgrounds, our predictions are realistic.  

Our forecasts are geared at new flavor-dependent neutrino interactions that are introduced by gauging three different accidental global lepton-number symmetries of the Standard Model, generated by $L_e-L_\mu$, $L_e-L_\tau$, and $L_\mu-L_\tau$, that have received prior attention in other experimental settings~\cite{Joshipura:2003jh, Bandyopadhyay:2006uh,  Heeck:2010pg, Chatterjee:2015gta, Khatun:2018lzs, Wise:2018rnb, Bustamante:2018mzu, Coloma:2020gfv}.  We focus on them because they can be gauged anomaly-free, so the only new particle introduced is a neutral vector boson that mediates the interaction.  Its mass and coupling strength are a priori undetermined.  Gauging $L_e-L_\mu$ and $L_e-L_\tau$ introduces new neutrino-electron interactions.  Gauging $L_\mu-L_\tau$ introduces new neutrino-neutron interactions.  

Under these interactions, electrons and neutrons source a flavor-dependent potential that may affect neutrino oscillations.  We concentrate on ultra-light mediators, with masses below $10^{-10}$~eV.  They induce interactions whose range is ultra-long --- ranging from hundreds of meters to Gpc, depending on the mass --- so that neutrinos may experience the potential sourced by a large number of nearby and distant electrons and neutrons in the Earth, the Moon, the Sun, the Milky Way, and the cosmological matter distribution~\cite{Bustamante:2018mzu}.  Yet, because the coupling strength may be tiny, their effects on the oscillation probabilities may be subtle and, therefore, testable with future experiments, like DUNE and T2HK.  

Our forecasts reveal two novel, promising perspectives.  First, while DUNE and T2HK, individually, should be able to improve on present-day upper limits on the coupling strength of the new interaction, their individual sensitivities are hampered by degeneracies due to uncertainties in the mixing angle $\theta_{23}$, the CP-violating phase, $\dcp$, and the neutrino mass ordering.  Yet, DUNE and T2HK have complementary capabilities: while T2HK is especially well-suited to measure $\theta_{23}$ and $\dcp$, DUNE is especially well-suited to measure the neutrino mass ordering.  Thus, \textbf{\textit{combining DUNE and T2HK removes parameter degeneracies, which tightens the upper limits on long-range neutrino interactions.}}  Second, and more importantly, \textbf{\textit{neither DUNE nor T2HK, by itself, may discover subdominant long-range interactions, owing to parameter degeneracies, but their combination may}}.  Thus, our forecasts stress the need for combining measurements in DUNE and T2HK to probe long-range interactions.

More broadly, our results illustrate the known need for complementarity in the future long-baseline neutrino program, not only to measure the standard mixing parameters, but to search for new physics. For flavor-dependent long-range neutrino interactions, a future global fit to oscillation data is poised to deliver substantially improved limits or transformative discovery.
%
\subsection*{Acknowledgments}
%
We thank Pilar Coloma, Sudipta Das, Iv\'an Esteban, Shirley Li, Swapna Mahapatra, Ashish Narang, and Yu-Dai Tsai for their helpful discussions and crucial input. S.K.A. acknowledges the support from the Department of Atomic Energy (DAE), Govt.~of India, under the Project Identification no.~RIO 4001. S.K.A.~acknowledges the financial support from the Swarnajayanti Fellowship (sanction order no.~DST/SJF/PSA-05/2019-20) provided by the Department of Science and Technology (DST), Govt.~of India, and the Research Grant (sanction order no.~SB/SJF/2020-21/21) provided by the Science and Engineering Research Board (SERB), Govt.~of India, under the Swarnajayanti Fellowship project. S.K.A.~would like to thank the United States-India Educational Foundation (USIEF) 
for providing the financial support through the Fulbright-Nehru Academic and Professional Excellence Fellowship (Award no.~2710/F-N APE/2021). M.S.~acknowledges the financial support from the DST, Govt.~of India (DST/INSPIRE Fellowship/2018/IF180059). M.B.~is supported by the {\sc Villum Fonden} under the project no.~29388. The numerical simulations are carried out using the ``SAMKHYA: High-Performance Computing Facility" at the Institute of Physics, Bhubaneswar, India.


\appendix


\section{Effects of long-range interactions on neutrino oscillation parameters}
\label{app:evol_mix_param} 

\renewcommand\thefigure{A\arabic{figure}}
\renewcommand\theHfigure{A\arabic{figure}}
\renewcommand\thetable{A\arabic{table}}
\renewcommand\theHtable{A\arabic{table}}
\setcounter{table}{0} 
\setcounter{figure}{0} 
\setcounter{equation}{0}

\begin{figure}[t!]
 \includegraphics[width=\linewidth]{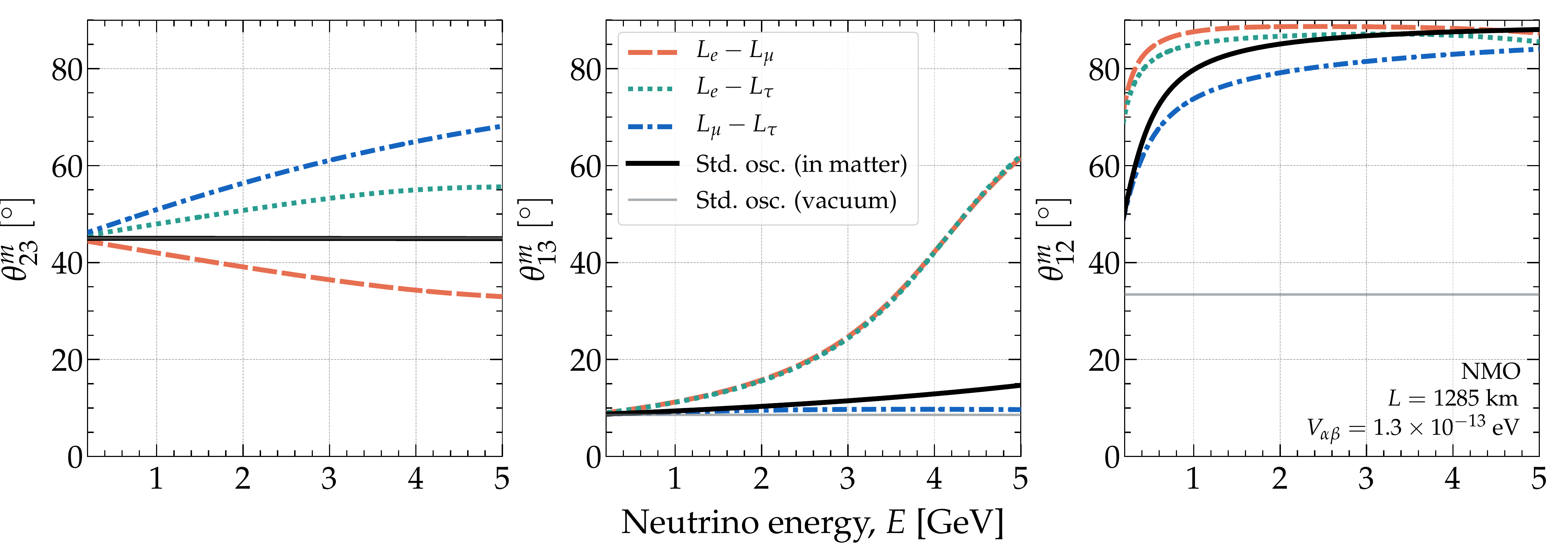}
 \caption{\textbf{\textit{Variation of the effective neutrino mixing angles with neutrino energy, for the three lepton-number symmetries, $L_\alpha-L_\beta$.}}  For this plot, we adopt the baseline, average matter potential, and approximate energy range of DUNE; see Section~\ref{sec:experiments_overview_dune}.
 For all the symmetries, we adopt an illustrative value of the new matter potential of $V_{\alpha\beta} = 1.3 \times 10^{-13}$~eV.  For comparison, we include results using only standard matter effects and in a vacuum.  The values of the mixing parameters in vacuum are from Table~\ref{tab:mix_param_benchmark}, except with $\sin^{2}\theta_{23} = 0.5$. See Appendix~\ref{app:evol_mix_param} for details.}
 \label{fig:evol_mix_param}
\end{figure}

\begin{figure}[t!]
 \includegraphics[width=\linewidth]{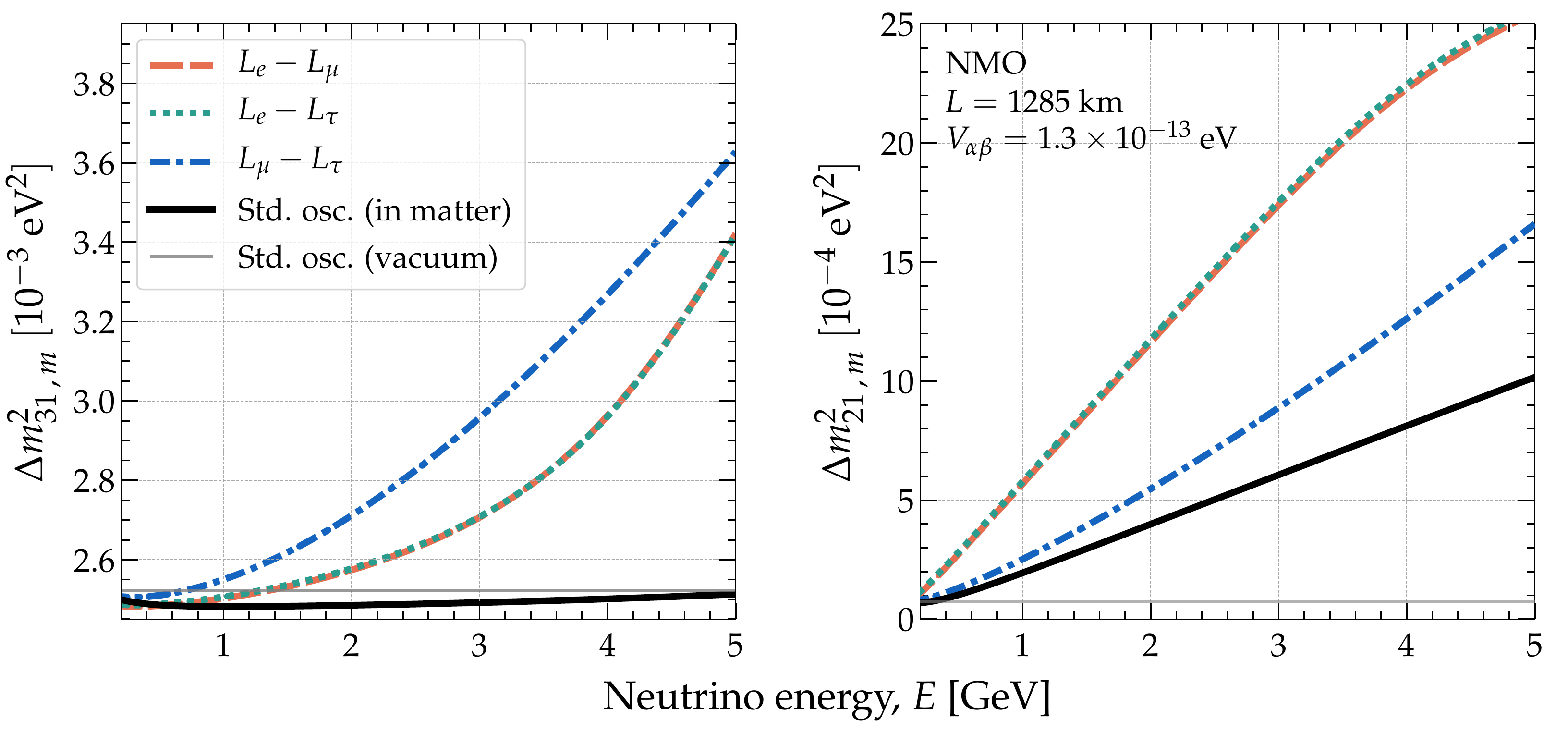}
 \caption{
 \textbf{\textit{Variation of the effective neutrino mass splittings with neutrino energy, for the three lepton-number symmetries, $L_\alpha-L_\beta$.}}  Same as \figu{evol_mix_param}, but for the mass splittings.  See Appendix~\ref{app:evol_mix_param} for details.}
 \label{fig:evol_mass_square}
\end{figure}

Figures~\ref{fig:evol_mix_param} and \ref{fig:evol_mass_square} show the variation with neutrino energy of, respectively, the mixing angles and mass splittings under the new matter interactions for the three symmetries, and geared at DUNE for illustration.  Their behavior when geared at T2HK is similar.  

Regarding the mixing angles, their behavior in \figu{evol_mix_param} backs the explanation of the behavior of the oscillation probabilities in Section~\ref{sec:experiments_probabilities}, in agreement with \Refes~\cite{Chatterjee:2015gta, Khatun:2018lzs, Agarwalla:2021zfr}.

Regarding the mass splittings, the energy at which the first oscillation maximum occurs in the probabilities (\figu{probabilities}) is determined mostly by $\Delta m^2_{31,m}$.  Figure~\ref{fig:evol_mass_square} shows that, under $L_e-L_\mu$ and $L_e-L_\tau$, $\Delta m^2_{31,m}$ evolves similarly with energy; this explains why the oscillation maxima for these two symmetries occur at approximately the same energy.  The change with energy of $\Delta m^2_{21,m}$ affects the $\nu_\mu \to \nu_\mu$ and $\nu_\mu \to \nu_\tau$ probabilities, but only for baselines of around 10000~km, as shown in \Refe~\cite{Agarwalla:2015cta}.


\section{Effects of oscillation parameter uncertainties on constraints}
\label{Appendix:2}

\renewcommand\thefigure{B\arabic{figure}}
\renewcommand\theHfigure{B\arabic{figure}}
\renewcommand\thetable{B\arabic{table}}
\renewcommand\theHtable{B\arabic{table}}
\setcounter{table}{0} 
\setcounter{figure}{0} 
\setcounter{equation}{0}

\begin{table}[t!]
 \centering
 \small
 \begin{tabular}{|c|c|c|c|c|c|c|c|c|c|}
  \hline 
  \multirow{3}{*}{Parameters in test event spectra} &
  \multicolumn{9}{c|}{Test-statistic, $\Delta \chi^2$} \\
  & \multicolumn{3}{c|}{$L_e-L_\mu$} & \multicolumn{3}{c|}{$L_e-L_\tau$} & \multicolumn{3}{c|}{$L_\mu-L_\tau$} \\
  & D & T & D+T & D & T & D+T & D & T & D+T \\   
  \hline
  Fixed $\dcp$, $\sin^2 \theta_{23}$, $\lvert \Delta m_{31}^2 \rvert$, $o$ & 23 & 8 & 25 & 37 & 10 & 41 & 50 & 18 & 58\\
  \hline
  Profiled over $\dcp$ only & 22 & 7 & 25 & 20 & 3 & 31 & 44 & 16 & 54 \\
  \hline
  Profiled over $\sin^2 \theta_{23}$ only & 10 & 2 & 19 & 35 & 8 & 40 & 25 & 8 & 48 \\
  \hline
  Profiled over $\lvert \Delta m_{31}^2 \rvert$ and $o$ only & 23 & 5 & 25 & 37 & 7 & 41 & 50 & 15 & 58 \\
  \hline
  \textbf{Profiled over $\dcp$, $\sin^2 \theta_{23}$, $\lvert \Delta m_{31}^2 \rvert$, $o$} & \textbf{6} & \textbf{2} & \textbf{19} & \textbf{18} & \textbf{3} & \textbf{31} & \textbf{11} & \textbf{7} & \textbf{44}\\
  \hline
 \end{tabular} 
 \caption{\textbf{\textit{Effect of profiling over the different parameters on the test-statistic for constraints.}}  The test-statistic is computed using \equ{delta_chi2_dune}, assuming $V_{\alpha\beta}^{\rm true} = 0$.  In this table, we report the values of the test-statistic calculated at an illustrative choice of the potential, $V_{\alpha\beta} = 3 \times 10^{-14}$~eV.  We show effects separately for DUNE (D), T2HK (T), and their combination (D+T).  Our main results are obtained by profiling over $\dcp$, $\sin^2 \theta_{23}$, $\lvert \Delta m_{31}^2\lvert$, and the neutrino mass ordering, $o$, \ie, the sign of $\Delta m_{31}^2$, in our prescription (see Section~\ref{sec:results_stat_methods}).  Results for partial profiling are shown only for the purpose of singling out the effect of different parameters.  The values of the parameters that are not profiled over are fixed to the benchmark values in Table~\ref{tab:mix_param_benchmark}.} 
 \label{table:A1}
\end{table}

Table~\ref{table:A1} illustrates how the uncertainty on different oscillation parameters affect the constraint the test-statistic, $\Delta \chi^2$, \ie, \equ{delta_chi2_dune} and similar ones, via profiling over them.  The table reports the test-statistic computed assuming that the true value of the potential is $V_{\alpha\beta}^{\rm true} = 0$, and evaluated at an illustrative test value of $V_{\alpha\beta} = 3 \times 10^{-14}$~eV.  The observations we make regarding Table~\ref{table:A1} hold also for other test values of $V_{\alpha\beta}$.  Our main results are computed by profiling the test-statistic over $\dcp$, $\sin^2\theta_{23}$, $\lvert \Delta m_{31}^2 \rvert$, and $o$.  Table~\ref{table:A1} shows what the effect is of profiling only over one or two of them at a time.

Because DUNE can measure $\lvert \Delta m^{2}_{31} \rvert$ and the mass ordering precisely, profiling over their values has little effect compared to keeping them fixed.  Similarly, because T2HK can measure $\theta_{23}$ and $\dcp$ precisely, profiling over their values has little effect compared to keeping them fixed.  Combining DUNE and T2HK affords both.   

Profiling over $\sin^2 \theta_{23}$ and $\dcp$ has the largest effect on the test-statistic.  The shrinking of $\Delta \chi^2$ by profiling over $\sin^2 \theta_{23}$, compared to keeping it fixed, comes via the $\nu_\mu \to \nu_\mu$ and $\bar{\nu}_\mu \to \bar{\nu}_\mu$ disappearance probabilities, which are $\propto \sin^2 \theta_{23}$ (see, \eg, Eq.~(33) in \Refe~\cite{Akhmedov:2004ny}).  The shrinking of $\Delta \chi^2$ by profiling over $\dcp$ comes instead via the $\nu_\mu \to \nu_e$ and $\bar{\nu}_\mu \to \bar{\nu}_e$ appearance probabilities, which depend on $\dcp$.  Thus, under $L_e-L_\mu$, the test-statistic is driven by the uncertainty in $\sin^2 \theta_{23}$ and $\dcp$ --- via profiling over them; under $L_e-L_\tau$, it is driven mostly by $\dcp$; and, under $L_\mu-L_\tau$, it is driven mostly by $\sin^2 \theta_{23}$.


\section{Constraints assuming inverted mass ordering }
\label{app:imo}

\renewcommand\thefigure{C\arabic{figure}}
\renewcommand\theHfigure{C\arabic{figure}}
\renewcommand\thetable{C\arabic{table}}
\renewcommand\theHtable{C\arabic{table}}
\setcounter{table}{0} 
\setcounter{figure}{0} 
\setcounter{equation}{0}

\begin{table}[b!]
 \centering
 \begin{tabular}{|c|c|c|c|c|c|c|}
  \hline 
  \multirow{2}{*}{} & \multicolumn{6}{c|}{Standard mixing parameters (IMO)} \\
   & $\sin^2 \theta_{12}$ & $\sin^2\theta_{23}$ & $\sin^2 \theta_{13}$ &
  $\frac{\Delta m^2_{31}}{10^{-3}\,\text{eV}^2}$  & $\frac{\Delta m^2_{21}}{10^{-5}\,\text{eV}^2}$  & $\delta_{\rm CP}\, (^\circ)$\\[0.8ex]
  \hline
  Benchmark & 0.303 & 0.569 & 0.0223 & 2.418 & 7.36 & 274  \\
  Status in fits & Fixed & Minimized & Fixed & Minimized & Fixed & Minimized \\
  Range & -- & [0.4, 0.6] & -- & [2.341, 2.501] & -- & [193, 342]\\
  \hline 
 \end{tabular}
 \caption{\textbf{\textit{Values of the standard mixing parameters used in our analysis, assuming that the true neutrino mass ordering is inverted.}}  Same as Table~\ref{tab:mix_param_benchmark}, made assuming that the true mass ordering is normal, but instead assuming that it is inverted (IMO).  The benchmark values are the best-fit values from \Refe~\cite{Capozzi:2021fjo}.}
 \label{tab:mix_param_benchmark_imo}
\end{table} 

\begin{figure}[t!]
 \centering
 \includegraphics[width=\linewidth]{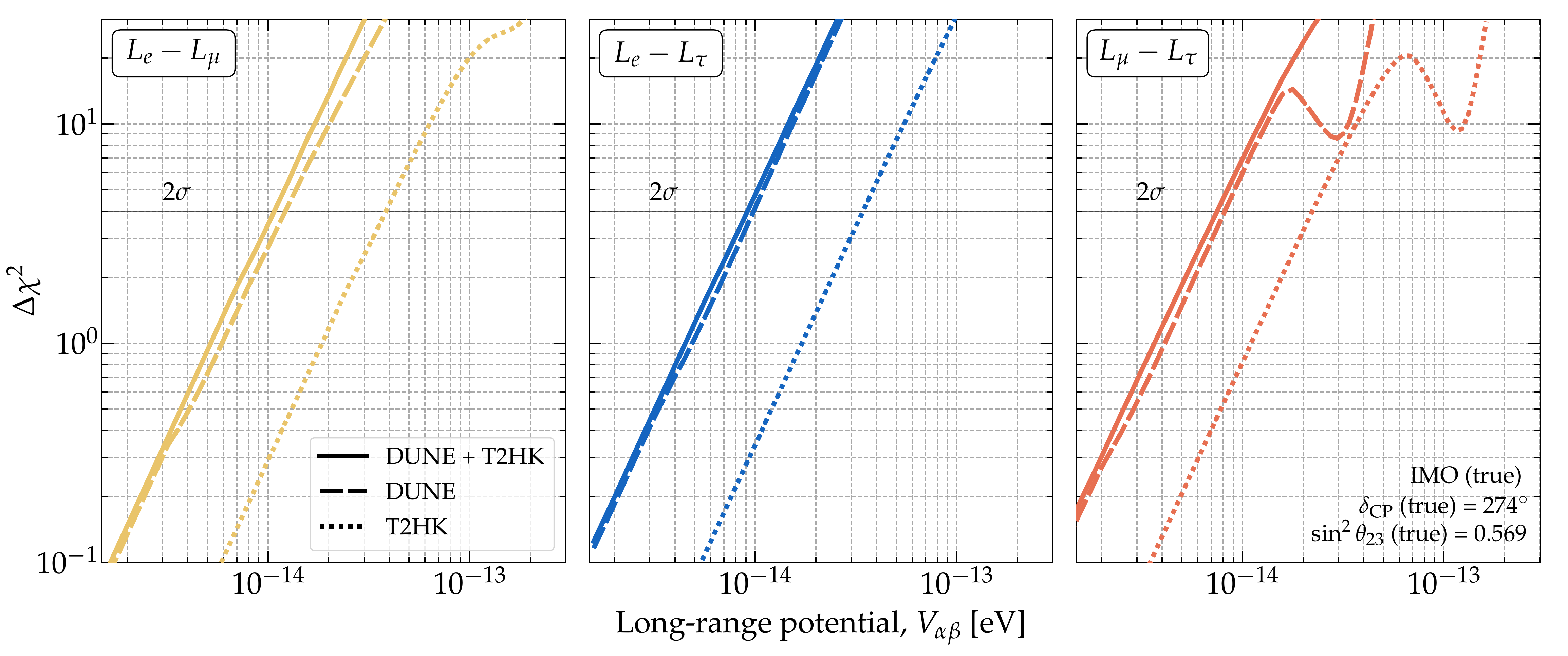}
 \caption{\textbf{\textit{Projected test-statistic used to constrain the long-range matter potentials $V_{e\mu}$, $V_{e\tau}$, and $V_{\mu\tau}$, using DUNE, T2HK, and their combination, assuming that the true neutrino mass ordering is inverted.}}  Same as \figu{test_statistics}, made assuming that the true mass ordering is normal, but instead assuming that it is inverted.  See Table~\ref{tab:upper_limits_potential_imo} for the resulting upper limits on the potentials.  See Sections~\ref{sec:results_stat_methods} and \ref{sec:results_constraints} in the main text, and Appendix~\ref{app:imo} for details.}
 \label{fig:test_statistics_imo}
\end{figure}

In the main text, we showed results obtained assuming that the true neutrino mass ordering is normal.  Here we show constraints  assuming instead that the true ordering is inverted.  They are broadly similar to the ones obtained under normal ordering (Section~\ref{sec:results_constraints}).  

Table~\ref{tab:mix_param_benchmark_imo} shows the values and allowed ranges of the mixing parameters that we use when assuming that the inverted mass ordering is true, taken from the global oscillation fit of \Refe~\cite{Capozzi:2021fjo}.  The values are similar as for normal ordering (Table~\ref{tab:upper_limits_potential}), with the important difference that for inverted ordering the benchmark value of $\theta_{23}$ lies in the higher octant, which has consequences for the sensitivity, as we point out below.

Figure~\ref{fig:test_statistics} shows how the test-statistics for constraints, \eg, \equ{delta_chi2_dune}, vary with the long-range potential, for the three symmetries and for DUNE, T2HK, and their combination, assuming the true mass ordering is inverted.  Their behavior is broadly similar to that in \figu{constraints_g_vs_m} for normal ordering.  However, in \figu{test_statistics_imo} the weakening of the sensitivity due to parameter degeneracies is milder than in \figu{test_statistics} because, assuming inverted mass ordering, $\theta_{23}$ lies in the higher octant, which lessens the influence of $\dcp$~\cite{Ballett:2016daj,Agarwalla:2022xdo}.

Table~\ref{tab:upper_limits_potential_imo} shows the resulting upper limits on the long-range potentials.  Like for normal ordering (Table~\ref{tab:upper_limits_potential}), they are strongest for $V_{\mu\tau}$.  Compared to normal ordering, the limits on $V_{e\mu}$ are stronger by a factor of 2 or more, and the limits on $V_{e\tau}$ are similarly weaker.

\begin{table}[t!]
\centering
 \begin{tabular}{ | c | *{4}{>{\centering\arraybackslash}p{2.25cm} |}}
 \hline
 \multirow{2}{*}{Detector} &
 \multicolumn{3}{c|}{Upper limit ($2\sigma$) on potential [$10^{-14}$~eV]} \\
  & $V_{e\mu}$ & $V_{e\tau}$ & $V_{\mu\tau}$ \\
 \hline
 DUNE        & $0.99$ & $3.4$ & $0.92$ \\
 T2HK        & $1.2$ & $3.9$ & $1.1$  \\
 DUNE + T2HK & $0.82$ & $2.2$ & $0.75$ \\
 \hline
 \end{tabular}
 \caption{\textbf{\textit{Projected upper limits (2$\sigma$) on the long-range matter potentials $V_{e\mu}$, $V_{e\tau}$, and $V_{\mu\tau}$, using DUNE, T2HK, and their combination, assuming that the true neutrino mass ordering is inverted.}}  Same as Table~\ref{tab:upper_limits_potential}, made assuming that the true mass ordering is normal, but instead assuming that it is inverted.  See \figu{test_statistics_imo} for the test-statistics from whence they originate and \figu{constraints_g_vs_m}. See Sections~\ref{sec:results_stat_methods} and \ref{sec:results_constraints}, and Appendix~\ref{app:imo} for details.}
 \label{tab:upper_limits_potential_imo}
\end{table}


\section{Normalization errors in event rates}
\label{Appendix:4}

\renewcommand\thefigure{D\arabic{figure}}
\renewcommand\theHfigure{D\arabic{figure}}
\renewcommand\thetable{D\arabic{table}}
\renewcommand\theHtable{D\arabic{table}}
\setcounter{table}{0} 
\setcounter{figure}{0} 
\setcounter{equation}{0}

\begin{table}[t!]
 \resizebox{\columnwidth}{!}{%
  \centering
  \begin{tabular}{|c|c|c|c|c|c|c|c|c|}
   \hline
   \multirow{3}{*}{Experiment}  & \multicolumn{8}{c|}{Normalization errors~[\%]}  \\
   & \multicolumn{4}{c|}{Signal, $\pi_{e,c}^s$} & \multicolumn{4}{c|}{Background, $\pi_{e,c,k}^b$}\\
   &  App.~$\nu$ & App.~$\bar{\nu}$ & Disapp.~$\nu$ & Disapp.~$\bar{\nu}$ & $\nu_{e}$, $\bar{\nu}_{e}$ CC & $\nu_{\mu}$, $\bar{\nu}_{\mu}$ CC & $\nu_{\tau}$, $\bar{\nu}_{\tau}$ CC & NC \\ 
   \hline
   DUNE & 2 & 2 & 5 & 5 & 5 & 5 & 20 & 10\\
   T2HK & 5 & 5 & 3.5 & 3.5 & 10 & 10 & -- & 10\\
   \hline
  \end{tabular}}
 \caption{\textbf{\textit{Normalization errors of the event rates associated to the signal and background detection channels in DUNE and T2HK.}}  They are shown separately for the neutrino ($\nu$) and antineutrino ($\bar{\nu}$) modes, and for the appearance (``App.'') and disappearance (``Disapp.'') channels.  Normalization errors are used to compute test event spectra, \equ{num_test}.  The values are taken from \Refe~\cite{Hyper-Kamiokande:2016srs, DUNE:2021cuw}.  See Section~\ref{sec:results_stat_methods} for details.}
 \label{tab:normalization_err}
\end{table}

Table~\ref{tab:normalization_err} shows the systematic normalization errors on the signal, $\pi_{e,c}^s$ and background event rates, $\pi_{e,c,k}^b$, of DUNE~\cite{DUNE:2021cuw} and T2HK~\cite{Hyper-Kamiokande:2016srs}.  In Section~\ref{sec:results_stat_methods}, we use them to compute test event spectra, \equ{num_test}, as part of computing the test-statistic.

\bibliographystyle{JHEP}
\bibliography{refer-lri}

\end{document}